\documentclass[reprint,twocolumn,aps,prb,showpacs]{revtex4-1}

\usepackage{amsmath}
\usepackage{amssymb}
\usepackage{graphicx}
\usepackage{dcolumn}
\usepackage{bm}
\usepackage[colorlinks=true, allcolors=blue]{hyperref}
\usepackage{epstopdf}
\usepackage[utf8]{inputenc}
\usepackage{amssymb}

\begin{document}

\title{Low-temperature T-linear resistivity in the strange-metal phase of overdoped cuprate
superconductors due to umklapp scattering from a spin excitation}

\author{Xingyu Ma}
\thanks{These authors contributed equally to this work}
\author{Minghuan Zeng}
\thanks{These authors contributed equally to this work}
\author{Zhangkai Cao}
\author{Shiping Feng}
\thanks{Corresponding author. E-mail: spfeng@bnu.edu.cn}

\affiliation{Department of Physics, Beijing Normal University, Beijing 100875, China}


\begin{abstract}
The strange-metal phase of overdoped cuprate superconductors exhibits a linear in
temperature resistivity in low temperatures, however, the origin of this remarkable anomaly
is still not well understood. Here the linear temperature dependence of the resistivity in
the strange-metal phase of overdoped cuprate superconductors is investigated. The momentum
dependence of the transport scattering rate is arisen from the electron umklapp scattering
mediated by the spin excitation, and is employed to calculate the resistivity by making use
of the Boltzmann equation. It is shown that the resistivity is mainly dominated by the
antinodal and nodal umklapp scattering. In particular, a very low temperature
$T_{\rm scale}$ scales with $\Delta^{2}_{p}$, where $\Delta_{p}$ is the minimal umklapp
vector at the antinode. In the low temperature above $T_{\rm scale}$, the resistivity is
linear in temperature with a coefficient that decreases with the increase of doping,
however, in the far lower temperature below $T_{\rm scale}$, the resistivity is instead
quadratic in temperature. The theory also shows that the same spin excitation that acts
like a bosonic glue to hold the electron pairs together also mediates scattering of
electrons in the strange-metal phase responsible for the linear in temperature resistivity
in low temperatures.
\end{abstract}

\pacs{74.25.Fy, 74.25.Nf, 74.20.Mn, 74.72.-h}

\maketitle

\section{Introduction}\label{Introduction}

The parent compound of cuprate superconductors is identified as a Mott insulator
\cite{Fujita12}, in which the absence of the electronic conduction is due to the strong
electron correlation. Superconductivity then emerges when charge carriers are doped into
this Mott insulator\cite{Bednorz86}, therefore the physical properties of cuprate
superconductors mainly depend on the extent of doping
\cite{Vishik18,Campuzano04,Damascelli03,Fink07,Keimer15,Hussey08,Timusk99,Kastner98}, and
the regimes have been classified into the underdoped, optimally doped, and overdoped,
respectively. At the temperature above the superconducting (SC) transition temperature
$T_{\rm c}$, the electron is in a normal-state. Although the same strong electron
correlation that leads to the Mott insulating state persists into the doped regime, the
normal-state retains a metallic character\cite{Bednorz86}.

In the underdoped regime, the normal-state is dominated by a pseudogap. This pseudogap
state is characterized by a substantial suppression of the density of the low-energy
excitations\cite{Vishik18,Campuzano04,Damascelli03,Fink07}. However, the resistivity in
the pseudogap phase is quadratic in temperature (T-quadratic)
\cite{Keimer15,Hussey08,Timusk99,Kastner98,Ando04,Barisic13,Pelc20}, as would be expected
from the standard Landau Fermi-liquid theory\cite{Schrieffer64,Abrikosov88,Mahan81}. On the
other hand, in the optimally doped and overdoped regimes, the normal-state is characterized
by a number of the anomalous low-temperature properties
\cite{Keimer15,Hussey08,Timusk99,Kastner98} in the sense that they do not fit in with the
standard Landau-Fermi liquid theory\cite{Schrieffer64,Abrikosov88,Mahan81}. This is why in
the optimally doped and overdoped regimes, the phase above $T_{\rm c}$ is so-called as
{\it the strange-metal phase} \cite{Keimer15}. In particular, in the early experimental
measurements \cite{Allen89,Gurvitch87,Takagi92}, it was observed that the variation of the
resistivity near the optimal doping is linear with temperature\cite{Takagi92}, extending
to low temperatures of a few kelvin and extrapolating to zero resistivity at zero
temperature. This remarkable behaviour of the linear in temperature (T-linear) resistivity
is in a striking contrast to the behaviour in conventional metals
\cite{Schrieffer64,Abrikosov88,Mahan81}, where the low-temperature resistivity follows one
of several simple power laws, and if the electron-electron scattering dominates, then the
resistivity decreases quadratically as the temperature decreases to zero. In the latter,
this low-temperature T-linear resistivity was detected experimentally in a wide doping
range of the overdoped regime
\cite{Martin90,Mandrus92,Ando01,Daou09,Cooper09,Legros19,Yuan22}. In particular, the
suppression of superconductivity with a magnetic field reveals that the low-temperature
T-linear resistivity persists down essentially to the zero temperature limit\cite{Ayres21}.
Recently, the systematic experimental observations in the heavily overdoped regime yielded
the low-temperature T-linear resistivity all the way up to the edge of the SC dome
\cite{Legros19,Yuan22,Ayres21,Grisso21}. After intensive investigations over more than
three decades, it has now become clear that the long-standing low-temperature T-linear
resistivity
\cite{Allen89,Gurvitch87,Takagi92,Martin90,Mandrus92,Ando01,Daou09,Cooper09,Legros19,Yuan22,Ayres21,Grisso21}
is a generic feature in the strange-metal phase of overdoped cuprate superconductors. In
this case, a key question posed by these experimental observations is raised: is there a
common bosonic excitation that is responsible for pairing the electrons also dominantly
scatters the electrons in the strange-metal phase responsible for the low-temperature
T-linear resistivity?

Although the low-temperature T-linear resistivity in the strange-metal phase of overdoped
cuprate superconductors is well established by now
\cite{Allen89,Gurvitch87,Takagi92,Martin90,Mandrus92,Ando01,Daou09,Cooper09,Legros19,Yuan22,Ayres21,Grisso21},
its origin remains the subject of the active research and debate. Theoretically, several
scenarios have been proposed for the origin of the T-linear resistivity
\cite{Varma89,Varma16,Varma20,Damle97,Sachdev11,Zaanen04,Luca07,Haldane18,Zaanen19,Hartnoll22,Hussey03,Rice17,Lee21}.
In particular, in the marginal Fermi-liquid phenomenology\cite{Varma89,Varma16,Varma20},
a single T-linear scattering rate is introduced responsible for the T-linear resistivity.
Moreover, it has been postulated that the T-linear behaviour can be attributed to the
strongly interacting critical state anchored at a quantum critical point (QCP) occurring
at doping of about $0.20$ where a phase transition is tuned to zero temperature
\cite{Varma20,Damle97,Sachdev11}. With the close relation to the physics of QCP, the
T-linear resistivity has been interpreted in terms of the Planckian
dissipation\cite{Zaanen04,Luca07,Haldane18,Zaanen19,Hartnoll22}
in which the relaxation-time achieves a putative universal minimum value, irrespective of
the underlying mechanisms. On the other hand, it has been argued that the elastic umklapp
scattering processes, which directly transfer momentum between the electron sea and the
underlying square lattice, lead to the T-linear resistivity in the strange-metal phase
\cite{Hussey03,Rice17}. More specifically, it has been shown recently that the resistance
arises from the electron umklapp scattering mediated by a critical bosonic mode\cite{Lee21},
where the resistivity is characterized by a highly anisotropic scattering rate. This highly
anisotropic scattering rate is T-linear near the umklapp point and becomes T-quadratic as
one moves away from the umklapp point, which therefore leads to a T-linear resistivity in
the low-temperature region and T-quadratic resistivity in the far-lower-temperature
region\cite{Lee21}. These studies\cite{Hussey03,Rice17,Lee21} and
many others\cite{Honerkamp01,Hartnoll12,Tabis21} therefore indicate that the electron
umklapp scattering dominates the momentum-relaxation mechanism of the electrical transport.
However, up to now, the origin of the low-temperature T-linear resistivity has not been
discussed starting from a microscopic theory, and no explicit calculations of the doping
dependence of the low-temperature T-linear resistivity has been made so far.
Superconductivity with the highest $T_{\rm c}$ emerges directly as an instability of the
strange-metal phase, and it thus has long been recognized that the understanding of the
essential physics of the strange-metal phase is crucial for the understanding of the mystery
of the unconventional superconductivity.

In the recent works within the framework of the kinetic-energy-driven superconductivity,
we have studied the low-energy electronic structure of cuprate superconductors both in the
SC-state\cite{Liu21,Cao21,Zeng22} and the strange-metal phase\cite{Feng16}, where the
electron normal self-energy in the particle-hole channel and electron anomalous self-energy
in the particle-particle channel are generated by the coupling of the electrons with the
spin excitations. In particular, the electrons are renormalized by the electron normal
self-energy, and then all the exotic features of the low-energy electronic structure arise
from this renormalization of the electrons\cite{Liu21,Cao21,Zeng22,Feng16}. In this paper,
we start from this low-energy electronic structure in the strange-metal phase of overdoped
cuprate superconductors\cite{Feng16} to study the nature of the doping dependence of the
low-temperature resistivity, where the momentum dependence of the transport scattering rate
is arisen from the electron umklapp scattering mediated by the same spin excitation, and is
employed to calculate the resistivity in terms of the Boltzmann equation. Our results show
that the momentum dependence of the transport scattering rate presents a similar behavior
of the single-particle scattering rate, and is largest at around the antinodes and smallest
at around the tips of the Fermi arcs, indicating that the resistivity is mainly dominated
by the antinodal and nodal umklapp scattering. In particular, a very low temperature
$T_{\rm scale}$ scales with $\Delta^{2}_{p}$, where $\Delta_{p}$ is the minimal umklapp
vector at the antinode. In the low-temperature region ($T>T_{\rm scale}$), the transport
scattering rate is T-linear with the coefficient that decreases with the increase of doping.
However, in the far-lower-temperature region ($T<T_{\rm scale}$), the transport scattering
rate is instead T-quadratic. This T-linear behaviour of the transport scattering rate in
the low-temperature region and the T-quadratic behaviour in the far-lower-temperature
region in turn generate respectively the T-linear resistivity in the low-temperature region
and the T-quadratic resistivity in the far-lower-temperature region. Our results therefore
indicate that the same spin excitation that is responsible for pairing the electrons also
mediates the electron umklapp scattering in the strange-metal phase responsible for the
low-temperature T-linear resistivity.

The rest of this paper is organized as follows. In Section \ref{Formalism}, we begin by a
short summary of the unconventional features of the low-energy electronic structure due to
the coupling of the electrons with the spin excitations, and then within the framework of
the Boltzmann transport theory, we formulate the essential aspects of the electron umklapp
scattering between a circular electron Fermi surface (EFS) and its umklapp partner mediated
by the same spin excitation for deriving the resistivity. In
Section \ref{electron-resistivity}, the Boltzmann equation is employed to study the doping
dependence of the low-temperature resistivity, where we show that both the
strengths of the nodal and antinodal umklapp scattering decrease with the decrease of
temperature. Finally, we give a summary and discussion in
Section \ref{summary}. In the Appendix \ref{electron-electron-collision}, we present the
details of the derivation of the electron-electron collision term in the Boltzmann equation.
In the Appendix \ref{reduced-kernel-function}, we present the details for the estimate of
the energy scale in the transport scattering rate at the umklapp point.

\section{Theory}\label{Formalism}

\subsection{$t$-$J$ model in the fermion-spin representation} \label{model-constraint}

The crystal structure of cuprate superconductors is characterized by the square-lattice
copper-oxide layers\cite{Fujita12,Bednorz86,Vishik18,Campuzano04,Damascelli03,Fink07},
which are sometimes considered to contain all the essential physics
\cite{Anderson87,Yu92,Lee06,Edegger07,Spalek22}. Immediately after the discovery of
superconductivity in cuprate superconductors, it was suggested that the fundamental
properties of the doped copper-oxide layer are properly accounted by the square-lattice
$t$-$J$ model\cite{Anderson87},
\begin{eqnarray}\label{tJ-model}
H=-\sum_{ll'\sigma}t_{ll'}C^{\dagger}_{l\sigma}C_{l'\sigma}
+\mu\sum_{l\sigma}C^{\dagger}_{l\sigma}
C_{l\sigma}+J\sum_{\langle ll'\rangle}{\bf S}_{l}\cdot {\bf S}_{l'},~~~~
\end{eqnarray}
acting on the restricted Hilbert-space with no double electron occupancy
$\sum_{\sigma}C^{\dagger}_{l\sigma}C_{l\sigma}\le1$, where the operator
$C^{\dagger}_{l\sigma}$ ($C_{l\sigma}$) denotes the creation (annihilation) operator of an
electron on site $l$ with spin $\sigma$, ${\bf S}_{l}$ is the spin operator of the electron
with its components $S_{l}^{x}$, $S_{l}^{y}$, and $S_{l}^{z}$, and $\mu$ is the chemical
potential. In this paper, the hopping of the constrained electrons $t_{ll'}$ is restricted
to the nearest-neighbor (NN) sites $\hat{\eta}$ with the hoping amplitude $t_{ll'}=t$ and
next NN sites $\hat{\tau}$ with the hoping amplitude $t_{ll'}=-t'$. The summation
$\langle ll'\rangle$ indicates a sum over the NN pairs, while the summation $ll'$ is taken
over all the NN and next NN pairs. Throughout this paper, we choose the parameters as
$t/J=2.5$ and $t'/t=0.3$ as in the previous discussions\cite{Feng16}. The magnitude of $J$
and the lattice constant of the square lattice are the energy and length units, respectively.
However, when necessary to compare with the experimental data, we set $J=1000$ K.

The essence of the strongly correlated physics is reflected in the on-site local constraint
of no double electron occupancy \cite{Yu92,Lee06,Edegger07,Spalek22,Zhang93}. To avoid the
double electron occupancy, we employ the fermion-spin transformation for the parametrization
of the constrained electron operators $C_{l\uparrow}$ and $C_{l\downarrow}$ as
\cite{Feng0494,Feng15},
\begin{eqnarray}\label{CSSFS}
C_{l\uparrow}=h^{\dagger}_{l\uparrow}S^{-}_{l},~~~~
C_{l\downarrow}=h^{\dagger}_{l\downarrow}S^{+}_{l},
\end{eqnarray}
respectively, where the spin operator $S_{l}$ keeps track of the spin degree of freedom of
the constrained electron, while the spinful fermion operator
$h_{l\sigma}=e^{-i\Phi_{l\sigma}}h_{l}$ keeps track of the charge degree of freedom of the
constrained electron together with some effects of spin configuration rearrangements due
to the presence of the doped hole itself (charge carrier). The advantages of this
fermion-spin approach (\ref{CSSFS}) can be summarized as: (i) the on-site local constraint
of no double occupancy is satisfied in actual analyses; (ii) the charge carrier or spin
{\it itself} is $U(1)$ gauge invariant\cite{Feng0494,Feng15}, and then the collective
mode for the spin is real and can be interpreted as the spin excitation responsible for
the dynamical spin response, while the electron quasiparticle as a result of the
charge-spin recombination of a charge carrier and a localized spin is not affected by the
statistical $U(1)$ gauge fluctuation\cite{Feng0494,Feng15}, and is responsible for the
electronic-state properties. This is why the fermion-spin approach (\ref{CSSFS}) is an
efficient calculation scheme which can provide a good description of the anomalous
properties of cuprate superconductors\cite{Feng0494,Feng15}.

In this fermion-spin representation (\ref{CSSFS}), the original $t$-$J$ model
(\ref{tJ-model}) can be rewritten explicitly as,
\begin{eqnarray}\label{cssham}
H&=&\sum_{ll'\sigma}t_{ll'}(h^{\dagger}_{l'\uparrow}h_{l\uparrow}S^{+}_{l}S^{-}_{l'}+
h^{\dagger}_{l'\downarrow}h_{l\downarrow}S^{-}_{l}S^{+}_{l'})-\mu_{\rm h}\sum_{l\sigma}
h^{\dagger}_{l\sigma}h_{l\sigma}\nonumber\\
&+&J_{\rm eff}\sum_{\langle ll'\rangle}{\bf S}_{l}\cdot {\bf S}_{l'},
\end{eqnarray}
where $S^{-}_{l}=S^{\rm x}_{l}-iS^{\rm y}_{l}$ and $S^{+}_{l}=S^{\rm x}_{l}+iS^{\rm y}_{l}$
are the spin-lowering and spin-raising operators for the spin $S=1/2$, respectively,
$J_{{\rm eff}}=(1-\delta)^{2}J$,
$\delta=\langle h^{\dagger}_{l\sigma}h_{l\sigma}\rangle=\langle h^{\dagger}_{l}h_{l}\rangle$
is the charge-carrier doping concentration, and $\mu_{\rm h}$ is the charge-carrier chemical
potential. As a natural consequence, the kinetic-energy term in the $t$-$J$ model
(\ref{tJ-model}) has been transferred as the coupling between charge and spin degrees of
freedom of the constrained electron, which reflects a basic fact that even the kinetic
energy term in the $t$-$J$ model (\ref{tJ-model}) has the strong Coulombic contribution due
to the restriction of no double electron occupancy at an any given site, and therefore
governs the unconventional features of cuprate superconductors.

\subsection{Coupling of electrons to the strongly dispersive spin excitation}
\label{Effective-propagator}

Starting from the $t$-$J$ model (\ref{cssham}) in the fermion-spin representation, we
\cite{Feng15,Feng0306,Feng12,Feng15a} have developed the kinetic-energy-driven
superconductivity, where the charge-carrier attractive interaction comes from the strong
coupling between charge and spin degrees of freedom of the constrained electron in the
kinetic energy of the $t$-$J$ model (\ref{cssham}), and induces the d-wave charge-carrier
pairing state, while the d-wave electron pairs originated from the d-wave charge-carrier
pairs are due to the charge-spin recombination\cite{Feng15a}, and their condensation
reveals the d-wave SC-state. In a similar to other distinct mechanisms for the
spin-fluctuation driven pairing\cite{Scalapino86,Miyake86,Monthoux91,Monthoux07}, the
kinetic-energy-driven SC mechanism is purely electronic without phonon, since the bosonic
glue is identified into an electron pairing mechanism not involving the phonon, the
external degree of freedom, but the internal spin degree of freedom of the constrained
electron itself, indicating that the strong electron correlation favors superconductivity.
Moreover, the kinetic-energy-driven SC-state is controlled by both the SC gap and
single-particle coherence, which leads to that the doping dependence of $T_{\rm c}$ exhibits
a dome-like shape with the underdoped and overdoped regimes on each side of the optimal
doping, where $T_{\rm c}$ reaches its maximum\cite{Feng15,Feng0306,Feng12,Feng15a}. In the
kinetic-energy-driven superconductivity, the self-consistent equations\cite{Feng15a} that
are satisfied by the single-particle diagonal and off-diagonal propagators in the SC-state
are obtained in terms of the Eliashberg formalism\cite{Eliashberg60}, and when the SC gap
parameter $\bar{\Delta}=0$, these self-consistent equations in the SC-state are reduced in
the normal-state as\cite{Feng16},
\begin{eqnarray}
G({\bf k},\omega)=G^{(0)}({\bf k},\omega)+G^{(0)}({\bf k},\omega)
\Sigma_{\rm ph}({\bf k},\omega)G({\bf k},\omega), ~~~~\label{EDGF}
\end{eqnarray}
where $G^{(0)}({\bf k},\omega)$ is the single-particle (diagonal) propagator of the $t$-$J$
model (\ref{tJ-model}) in the tight-binding approximation, and has been derived as
$G^{(0)-1}({\bf k},\omega)=\omega-\varepsilon_{\bf k}$. In this case, the single-particle
propagator $G({\bf k},\omega)$ in Eq. (\ref{EDGF}) can be expressed explicitly as,
\begin{eqnarray}\label{EGF}
G({\bf k},\omega)={1\over \omega-\varepsilon_{\bf k}-\Sigma_{\rm ph}({\bf k},\omega)},
\end{eqnarray}
where $\varepsilon_{\bf k}=-4t\gamma_{\bf k}+4t'\gamma_{\bf k}'+\mu$ is the electron energy
dispersion in the tight-binding approximation,
\begin{figure}[h!]
\centering
\includegraphics[scale=0.7]{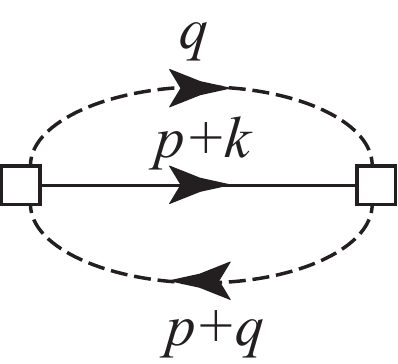}
\caption{The skeletal diagram for the electron normal self-energy for scattering electrons
from the strongly dispersive spin excitations. The solid-line represents the electron
propagator $G$, and the dashed-line depicts the spin propagator $D^{(0)}$, while $\square$
describes the bare vertex function $\Lambda$. \label{self-energy-diagram}}
\end{figure}
with
$\gamma_{\bf k}=({\rm cos}k_{x}+{\rm cos} k_{y})/2$ and
$\gamma_{\bf k}'={\rm cos}k_{x}{\rm cos}k_{y}$, while the electron normal self-energy
$\Sigma_{\rm ph}({\bf k},\omega)$ sketched in Fig. \ref{self-energy-diagram} has been
derived as\cite{Feng16},
\begin{eqnarray}\label{ESE}
\Sigma_{\rm ph}({\bf k},\omega) &=& -4\int^{\infty}_{-\infty}{{\rm d}\omega'\over 2\pi}
\int^{\infty}_{-\infty}{{\rm d}\omega''\over 2\pi}{n_{\rm B}(\omega'')+n_{\rm{F}}(\omega')
\over\omega''-\omega'+\omega}\nonumber\\
&\times& {t^{2}\over N}\sum_{\bf p} {\rm Im}G({\bf p}+{\bf k},\omega'){\rm Im}
P^{(0)}({\bf k},{\bf p},\omega''),~~~~
\end{eqnarray}
where $N$ is the number of lattice sites, $n_{\rm B}(\omega)$ and $n_{\rm F}(\omega)$ are
the boson and fermion distribution functions, respectively,
${\rm Im}P^{(0)}({\bf k},{\bf p},\omega)$ is the imaginary part of
$P^{(0)}({\bf k},{\bf p},\omega)$, while $P^{(0)}({\bf k},{\bf p},\omega)$ is so-called as
the effective spin propagator, which describes the nature of the spin excitation, and can
be expressed as,
\begin{eqnarray}\label{ESP-1}
P^{(0)}({\bf k},{\bf p},\omega)={1\over N}\sum_{\bf q}\Lambda^{2}_{{\bf p}+{\bf q}+{\bf k}}
\Pi({\bf p},{\bf q},\omega),
\end{eqnarray}
with the bare vertex function $\Lambda_{{\bf k}}=4\gamma_{\bf k}-4(t'/t)\gamma_{\bf k}'$
and the spin bubble $\Pi({\bf p},{\bf q},\omega)$. This spin bubble
$\Pi({\bf p},{\bf q},\omega)$ is a convolution of two spin propagators, and has been
evaluated as\cite{Feng16},
\begin{eqnarray}\label{spin-bubble-1}
\Pi({\bf p},{\bf q},ip_{m})={1\over\beta}\sum_{iq_{m}}D^{(0)}({\bf q},iq_{m})
D^{(0)}({\bf q}+{\bf p},iq_{m}+ip_{m}),\nonumber\\
\end{eqnarray}
where $p_{m}$ and $q_{m}$ are the bosonic Matsubara frequencies, while the spin propagator
$D^{(0)}({\bf k},\omega)$ in the mean-field (MF) level has been derived as,
\begin{eqnarray}\label{SGF-1}
D^{(0)}({\bf k},\omega)={B_{\bf k}\over\omega^{2}-\omega^{2}_{\bf k}},
\end{eqnarray}
with the MF spin excitation energy dispersion $\omega_{\bf k}$ and the weight function of
the spin excitation spectrum $B_{\bf k}$ that have been obtained explicitly as\cite{Feng15},
\begin{widetext}
\begin{subequations}\label{spin-function}
\begin{eqnarray}
\omega^{2}_{\bf k}&=&\alpha\lambda_{1}^{2}\left [{1\over 2}\epsilon\chi_{1}\left (A_{11}
-\gamma_{\bf k}\right)(\epsilon-\gamma_{\bf k})+\chi^{\rm z}_{1}\left (A_{12}-\epsilon
\gamma_{\bf k}\right )(1-\epsilon\gamma_{\bf k})\right ]+\alpha\lambda_{2}^{2}\left [
\left (\chi^{\rm z}_{2}\gamma_{\bf k}'-{3\over 8}\chi_{2}\right )\gamma_{\bf k}'+A_{13}
\right ]\nonumber\\
&+&\alpha\lambda_{1}\lambda_{2} \left [\chi^{\rm z}_{1}(1-\epsilon\gamma_{\bf k})
\gamma_{\bf k}'+{1\over 2}(\chi_{1}\gamma_{\bf k}'-C_{3})(\epsilon-\gamma_{\bf k})
+\gamma_{\bf k}'(C^{\rm z}_{3}-\epsilon\chi^{\rm z}_{2}\gamma_{\bf k})-{1\over 2}
\epsilon (C_{3}-\chi_{2}\gamma_{\bf k})\right ], ~~~~~~~\label{MFSES}\\
B_{\bf k}&=&\lambda_{1}[2\chi^{\rm z}_{1}(\epsilon\gamma_{\bf k}-1)
+\chi_{1}(\gamma_{\bf k}-\epsilon)]-\lambda_{2}(2\chi^{\rm z}_{2}\gamma_{\bf k}'
-\chi_{2}), \label{spin-weight-function}
\end{eqnarray}
\end{subequations}
\end{widetext}
where $\epsilon=1+2t\phi_{1}/J_{\rm eff}$, $\lambda_{1}=8J_{\rm eff}$,
$\lambda_{2}=16\phi_{2}t'$, the charge-carrier's particle-hole parameters
$\phi_{1}=\langle h^{\dagger}_{l\sigma}h_{l+\hat{\eta}\sigma}\rangle$ and
$\phi_{2}=\langle h^{\dagger}_{l\sigma}h_{l+\hat{\tau}\sigma}\rangle$,
$A_{11}=[(1-\alpha)/8-\alpha\chi^{\rm z}_{1}/2+\alpha C_{1}]/(\alpha\chi_{1})$,
$A_{12}=[(1-\alpha)/16-\alpha\epsilon\chi_{1}/8+\alpha C^{\rm z}_{1}]
/(\alpha\chi^{\rm z}_{1})$, $A_{13}=[(1-\alpha)/(8\alpha)-\chi^{\rm z}_{2}/2+C_{2}]/2$, the
spin correlation functions $\chi_{1}=\langle S_{l}^{+}S_{l+\hat{\eta}}^{-}\rangle$,
$\chi_{2}=\langle S_{l}^{+}S_{l+\hat{\tau}}^{-}\rangle$,
$\chi^{\rm z}_{1}=\langle S_{l}^{\rm z}S_{l+\hat{\eta}}^{\rm z} \rangle$,
$\chi^{\rm z}_{2}=\langle S_{l}^{\rm z}S_{l+\hat{\tau}}^{\rm z} \rangle$,
$C_{1}=(1/16)\sum_{\hat{\eta},\hat{\eta'}}\langle S_{l+\hat{\eta}}^{+}S_{l+\hat{\eta'}}^{-}
\rangle$,
$C^{\rm z}_{1}=(1/16)\sum_{\hat{\eta},\hat{\eta'}}\langle S_{l+\hat{\eta}}^{z}
S_{l+\hat{\eta'}}^{z}\rangle$, $C_{2}=(1/16)\sum_{\hat{\tau},\hat{\tau'}}\langle S_{l+\hat{\tau}}^{+}S_{l+\hat{\tau'}}^{-}\rangle$,
$C_{3}=(1/4) \sum_{\hat{\tau}}\langle S_{l+\hat{\eta}}^{+}S_{l+\hat{\tau}}^{-}\rangle$, and
$C^{\rm z}_{3}=(1/4)\sum_{\hat{\tau}}\langle S_{l+\hat{\eta}}^{\rm z}
S_{l+\hat{\tau}}^{\rm z}\rangle$.
In order to satisfy the sum rule of the correlation function
$\langle S^{+}_{l}S^{-}_{l}\rangle=1/2$ in the case without an antiferromagnetic long-range
order (AFLRO), the important decoupling parameter $\alpha$ has been introduced in the
calculation\cite{Feng15}, which can be regarded as the vertex correction. At the
half-filling, the degree of freedom is local spin only, where $\epsilon=1$,
$\lambda_{2}=0$, $\chi^{\rm z}_{1}=\chi_{1}/2$, $C^{\rm z}_{1}=C_{1}/2$, and then the above
MF spin excitation energy dispersion $\omega_{\bf k}$ and the weight function $B_{\bf k}$
in Eq. (\ref{spin-function}) are reduced as,
$\omega_{\bf k}=\lambda_{1}\sqrt{\alpha|\chi_{1}|(1-\gamma^{2}_{\bf k})}$ and
$B_{\bf k}=-2\lambda_{1}\chi_{1}(1-\gamma_{\bf k})$, respectively. This spin excitation
energy dispersion at the half-filling is the standard spin-wave dispersion. However, in
the doped regime without an AFLRO, although the expression form of the MF spin excitation
energy dispersion $\omega_{\bf k}$ in Eq. (\ref{MFSES}) is rather complicated, the
essential feature of the spin-wave nature is the same as that in the case of the
half-filling\cite{Feng15}.

Substituting the above MF spin propagator (\ref{SGF-1}) into Eq. (\ref{spin-bubble-1}),
the spin bubble $\Pi({\bf p},{\bf q},\omega)$ can be derived as,
\begin{eqnarray}\label{spin-bubble}
\Pi({\bf p},{\bf q},\omega)=-{\bar{W}^{(1)}_{{\bf p}{\bf q}}\over\omega^{2}
-[\omega^{(1)}_{{\bf p}{\bf q}}]^{2}}+{\bar{W}^{(2)}_{{\bf p}{\bf q}}\over\omega^{2}
-[\omega^{(2)}_{{\bf p}{\bf q}}]^{2}},~~~~
\end{eqnarray}
and then the effective spin propagator $P^{(0)}({\bf k},{\bf p},\omega)$ in
Eq. (\ref{ESP-1}) is obtained directly from the above spin bubble (\ref{spin-bubble}),
where the spin excitation energy dispersions
$\omega^{(1)}_{{\bf p}{\bf q}}=\omega_{{\bf q}+{\bf p}}+\omega_{\bf q}$ and
$\omega^{(2)}_{{\bf p}{\bf q}}=\omega_{{\bf q}+{\bf p}}-\omega_{\bf q}$, and the weight
functions of the effective spin excitation spectrum,
\begin{eqnarray}
\bar{W}^{(1)}_{{\bf p}{\bf q}}&=&{B_{\bf q}B_{{\bf q}+{\bf p}}\over 2\omega_{\bf q}
\omega_{{\bf q}+{\bf p}}}\omega^{(1)}_{{\bf p}{\bf q}}[n_{\rm B}(\omega_{{\bf q}+{\bf p}})
+n_{\rm B}(\omega_{\bf q})+1], ~~~~~
\end{eqnarray}
\begin{eqnarray}
\bar{W}^{(2)}_{{\bf p}{\bf q}}&=&{B_{\bf q}B_{{\bf q}+{\bf p}}\over 2\omega_{\bf q}
\omega_{{\bf q}+{\bf p}}}\omega^{(2)}_{{\bf p}{\bf q}}[n_{\rm B}(\omega_{{\bf q}+{\bf p}})
-n_{\rm B}(\omega_{\bf q})].
\end{eqnarray}
It thus shows that the essential behaviors of the spin excitation energy dispersions
$\omega^{(1)}_{{\bf p}{\bf q}}$ and $\omega^{(2)}_{{\bf p}{\bf q}}$ are mainly governed by
the essential behaviors of the MF spin excitation energy dispersion $\omega_{\bf k}$ in
Eq. (\ref{MFSES}).

With the above effective spin propagator (\ref{ESP-1}), the electron normal self-energy
$\Sigma_{\rm ph}({\bf k},\omega)$ in Eq. (\ref{ESE}) has been obtained explicitly in Ref.
\onlinecite{Feng16}. In particular, it should be emphasized that all the order parameters
and chemical potential $\mu$ in the calculation of $\Sigma_{\rm ph}({\bf k},\omega)$ have
been determined by the self-consistent calculation \cite{Feng15,Feng0306,Feng12,Feng15a}.
In this sense, our calculation for $\Sigma_{\rm ph}({\bf k},\omega)$ is controllable
without using adjustable parameters. Moreover, the sharp peaks visible for low-temperature
in $\Sigma_{\rm ph}({\bf k},\omega)$ and $P^{(0)}({\bf k},{\bf p},\omega)$ are actually a
$\delta$-function, broadened by a small damping used in the numerical calculation for a
finite lattice\cite{Brinckmann01,Restrepo23}. The calculation in this paper for
$\Sigma_{\rm ph}({\bf k},\omega)$ and $P^{(0)}({\bf k},{\bf p},\omega)$ is performed
numerically on a $160\times 160$ lattice in momentum space, with the infinitesimal
$i0_{+}\rightarrow i\Gamma$ replaced by a small damping $\Gamma=0.05J$.

\subsection{Electron Fermi surface}\label{Octet-model}

The single-particle spectrum function $A({\bf k},\omega)$ now can be obtained directly
from the above single-particle propagator ({\ref{EGF}}) as,
\begin{eqnarray}\label{ESF}
A({\bf k},\omega)&=&-2{\rm Im}G({\bf k},\omega)={2\Gamma_{\bf k}(\omega)\over
[\omega-\bar{E}_{\bf k}(\omega)]^{2}+\Gamma^{2}_{\bf k}(\omega)},~~~~~~
\end{eqnarray}
with the corresponding single-particle scattering rate $\Gamma_{\bf k}(\omega)$ and
renormalized band structure $\bar{E}_{\bf k}(\omega)$,
\begin{subequations}
\begin{eqnarray}
\Gamma_{\bf k}(\omega)&=& |{\rm Im}\Sigma_{\rm ph}({\bf k},\omega)|, ~~~~~
\label{SPSR}\\
\bar{E}_{\bf k}(\omega)&=& \varepsilon_{\bf k}+{\rm Re}\Sigma_{\rm ph}({\bf k},\omega),~~~~~
\label{RBS}
\end{eqnarray}
\end{subequations}
where ${\rm Re}\Sigma_{\rm ph}({\bf k},\omega)$ and ${\rm Im}\Sigma_{\rm ph}({\bf k},\omega)$
are the real and imaginary parts of the electron normal self-energy
$\Sigma_{\rm ph}({\bf k},\omega)$, respectively.

The shape of EFS has deep consequences for the low-energy electronic properties
\cite{Vishik18,Campuzano04,Damascelli03,Fink07,Keimer15}, and has been also central to
addressing electrical transport\cite{Hussey08,Timusk99,Kastner98}. In the previous studies
\cite{Feng16}, the topology of EFS in the strange-metal phase of cuprate superconductors
has been discussed in terms of the intensity map of the single-particle spectral function
(\ref{ESF}) at zero energy $\omega=0$, where it has been shown that (i) the EFS continuous
contour is determined by the poles of the single-particle propagator ({\ref{EGF}}) at zero
energy, i.e.,
$\varepsilon_{\bf k}+{\rm Re}\Sigma_{\rm ph}({\bf k},0)=\bar{\varepsilon}_{\bf k}=0$, with
the renormalized electron energy dispersion
$\bar{\varepsilon}_{\bf k}=Z_{\rm F}\varepsilon_{\bf k}$ and the single-particle coherent
weight $Z^{-1}_{\rm F}=1-{\rm Re}\Sigma_{\rm pho}({\bf k},0)\mid_{{\bf k}=[\pi,0]}$,
where $\Sigma_{\rm pho}({\bf k},\omega)$ is the antisymmetric part of the electron normal
self-energy $\Sigma_{\rm ph}({\bf k},\omega)$;
\begin{figure}[h!]
\centering
\includegraphics[scale=0.65]{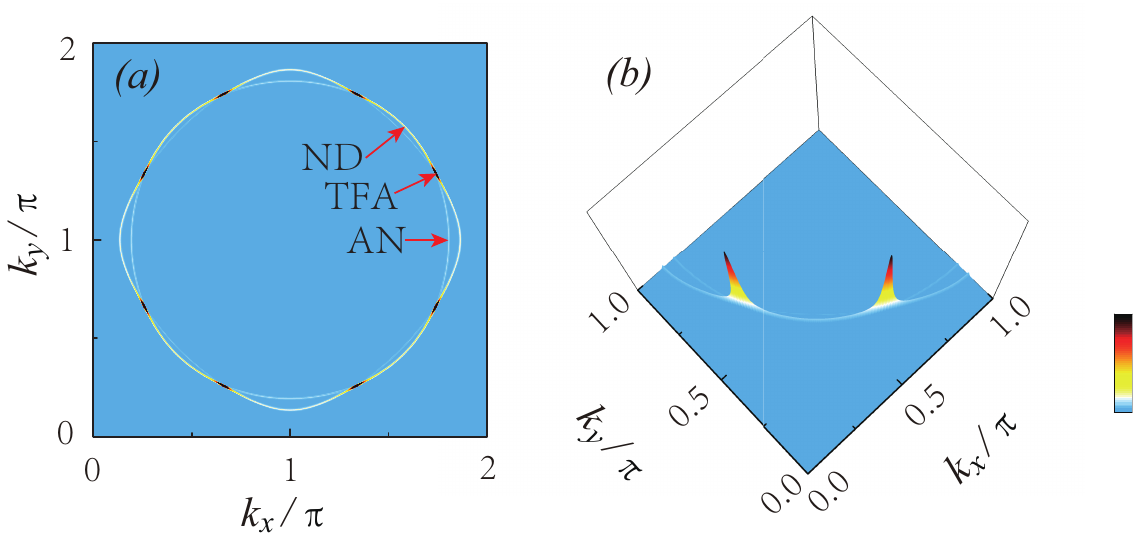}
\caption{(Color online) (a) The map of the electron Fermi surface and (b) the surface plot
of the electron spectral function for zero energy $\omega=0$ at $\delta=0.18$ with
$T=0.002J$, where the Brillouin zone center has been shifted by [$\pi,\pi$], and AN, TFA,
and ND denote the antinode, tip of the Fermi arc, and node, respectively. \label{EFS-map}}
\end{figure}
(ii) however, a strong redistribution of the
spectral weight on EFS is induced by the highly anisotropic momentum dependence of the
single-particle scattering rate $\Gamma_{\bf k}(\omega)$. For a convenience in the following
discussions of the electrical transport, we plot (a) the EFS map and (b) the direct surface
plot of the single-particle spectral function $A({\bf k},\omega)$ for zero energy $\omega=0$
at doping $\delta=0.18$ with temperature $T=0.002J$ in Fig. \ref{EFS-map}, where the
Brillouin zone (BZ) center has been shifted by [$\pi,\pi$], and AN, TFA, and ND indicate the
antinode, tip of the Fermi arc, and node, respectively. The most noteworthy in
Fig. \ref{EFS-map} are the following: (i) the spectral weight at around the antinodal region
is suppressed strongly, reflecting a basic fact that EFS at around the antinodal region can
not be observed experimentally
\cite{Norman98,Shi08,Sassa11,Meng11,Horio16,Loret18,Chen19,Loret17,Kaminski15,Comin14,Fujita14,Hufner08,He14};
(ii) the spectral weight at around the nodal region is suppressed modestly, leading to the
formation of the disconnected Fermi arcs
\cite{Norman98,Shi08,Sassa11,Meng11,Horio16,Loret18,Chen19,Loret17,Kaminski15,Comin14,Fujita14,Hufner08,He14},
where the Fermi arc increases its length as a function of doping
\cite{Chen19,Loret17,Kaminski15,Comin14,Fujita14,Hufner08,He14}, and then it evolves into a
continuous contour in momentum space near the end of the SC dome; (iii) however, almost all
the spectral weight on the Fermi arcs is assembled at around the tips of the Fermi arcs
\cite{Norman98,Shi08,Sassa11,Meng11,Horio16,Loret18,Chen19,Loret17,Kaminski15,Comin14,Fujita14,Hufner08,He14}.
In other words, the electrons at around the tips of the Fermi arcs have a largest density
of states, and then the low-energy electronic properties are largely governed by these
electrons at around the tips of the Fermi arcs. In particular, it has been observed
experimentally that these characteristic features shown in Fig. \ref{EFS-map} in the zero
energy case can persist into the case for a finite binding-energy even in the optimally and
overdoped regimes\cite{He14,Chatterjee06}. More importantly, the suppression of the spectral
weight at around the antinodal and nodal regions can affect the electrical transport in two
ways\cite{Timusk99}: through the reduction of the number of current-carrying states, and
secondly, through the reduction in the density of electron excitations at around the
antinodal and nodal regions.

\begin{figure}[h!]
\centering
\includegraphics[scale=0.75]{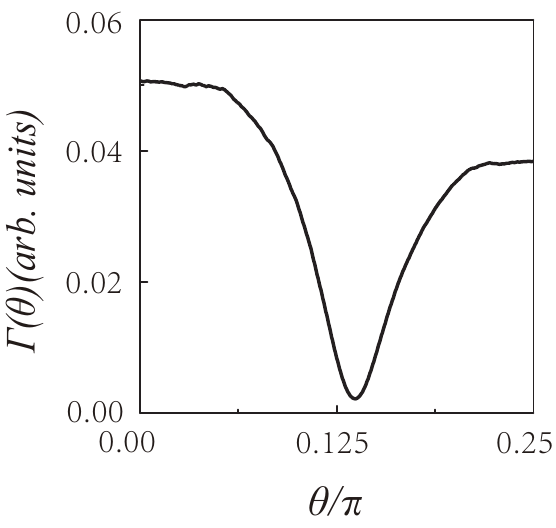}
\caption{The single-particle scattering rate $\Gamma(\theta)$ as a function of Fermi angle
$\theta$ at $\delta=0.18$ with $T=0.05J$ for $\omega=0$.
\label{single-particle-scattering}}
\end{figure}

In our previous discussions \cite{Feng16}, it has been shown that the origin of the spectral
redistribution to form the Fermi arcs can be attributed to the highly anisotropic momentum
dependence of the single-particle scattering rate $\Gamma(\theta)$, where $\Gamma(\theta)$
is defined as $\Gamma(\theta)=\Gamma_{{\rm k}_{\rm F}(\theta)}(0)$, and $\theta$ is the
Fermi angle. To see this highly anisotropic $\Gamma(\theta)$ in momentum space more clearly,
we plot the angular dependence of $\Gamma(\theta)$ along EFS from the antinode to the node
at $\delta=0.18$ with $T=0.05J$ in Fig. \ref{single-particle-scattering}, where the actual
minimum of $\Gamma(\theta)$ does not appear at around the nodal region, but resides exactly
at around the tip of the Fermi arc. However, the maximal $\Gamma(\theta)$ appears at around
the antinodal region, and then $\Gamma(\theta)$ decreases when the Fermi angle is moved
away from the antinode. In particular, $\Gamma(\theta)$ at around the nodal region is
smaller than that around the antinodal region. This angular dependence of $\Gamma(\theta)$
therefore leads to the spectral redistribution to form the Fermi arcs with almost all the
spectral weight inhabited at around the tips of the Fermi arcs.

With the help of the above single-particle spectrum function (\ref{ESF}), we have also
studied the renormalization of the electrons in the strange-metal phase of overdoped
cuprate superconductors\cite{Feng16}, including the EFS instability driven charge order
together with the related octet scattering model
\cite{Comin14,Fujita14,He14,Chatterjee06,Comin16,Yin21,Pan01,Fujita19}, the complicated
line-shape in the energy distribution curve\cite{Dessau91,Norman97,Campuzano99,Wei08,DMou17},
and the kink in the electron dispersion
\cite{Kaminski01,Zhou03,Anzai10,He13,Yang19,Dessau93},
and the obtained results are well consistent with the corresponding experimental
observations. In particular, it should be emphasized that the similar results
\cite{Eschrig00,Manske01,Eschrig02,Eschrig06} have been also obtained based on the
phenomenological spin-fermion approach \cite{Eschrig06,Monthoux91}, where the coupling of a
spin excitation to electron quasiparticles leads to the emergence of the spectral dip in the
energy distribution curve and the kink in the quasiparticle dispersion
\cite{Eschrig00,Manske01,Eschrig02,Eschrig06}. These results
\cite{Feng16,Eschrig00,Manske01,Eschrig02,Eschrig06}
therefore show that the same spin excitation that is responsible for pairing the electrons
also dominantly scatters the electrons in the strange-metal phase responsible for the
low-energy electronic structure.

\subsection{Boltzmann equation}\label{Boltzmann-theory}

Although the magnitude of the single-particle scattering rate $\Gamma(\theta)$ shown in
Fig. \ref{single-particle-scattering} at a given Fermi angle is different from that of the
corresponding transport scattering rate \cite{Varma20}, both the scattering rates may have
a similar behaviour of the angular dependence. In this sense, the result of the angular
dependence of $\Gamma(\theta)$ in Fig. \ref{single-particle-scattering} also indicates that
the important electron scattering responsible for the resistivity is mainly concentrated
at around the antinodes and nodes. For the discussions of the transport properties,
it is needed to determine how the momentum distribution relaxes in the vicinity
of these antinodes and nodes. However, in the strange-metal phase of overdoped cuprate
superconductors, there are no well-defined quasiparticle excitations
\cite{Keimer15,Hussey08,Timusk99,Kastner98}. In this case, the momentum distribution relaxes
in the system can be discussed in terms of the memory-matrix transport formalism
\cite{Mahajan13,Hartnoll14,Patel14,Lucas15,Vieira20,Mandal21} or by solving the Boltzmann
equation with the input of the scattering processes\cite{Abrikosov88,Mahan81}. The
memory-matrix transport formalism\cite{Mahajan13,Hartnoll14,Patel14,Lucas15,Vieira20,Mandal21}
has a crucial advantage of not relying on the existence of well-defined quasiparticle
excitations. In particular, this memory-matrix transport approach has been employed to study
the electrical transport in the strange-metal phase of different strongly correlated models
\cite{Mahajan13,Hartnoll14,Patel14,Lucas15,Vieira20,Mandal21}, and the obtained results are
consistent with the rather severe set by experiments. On the other hand, in the Boltzmann
transport theory\cite{Abrikosov88,Mahan81}, it is crucial to assume that either the
quasiparticle excitations are well-defined or that the effective interaction between
electrons via various bosonic modes can be treated within the Eliashberg approach
\cite{Eliashberg60} as shown by Prange and Kadanoff\cite{Prange64} for an electron-phonon
system. In this paper, we study the electrical transport in the strange-metal phase of
overdoped cuprate superconductors by solving the Boltzmann equation with the input of the
scattering process, since the electron interaction mediated by the spin excitation in the
framework of the kinetic-energy-driven superconductivity
\cite{Feng15,Feng0306,Feng12,Feng15a} is treated within the Eliashberg approach
\cite{Eliashberg60} as we have mentioned in subsection \ref{Effective-propagator}. In the
Boltzmann transport theory\cite{Abrikosov88,Mahan81}, the essential behaviour of the
electrons is depicted by the distribution function $f({\bf r},{\bf k},t)$. In the following
discussions, we focus on the dc conductivity in the homogeneous system only, where the
position and time dependence in the distribution function are absent, and then the
distribution function satisfies the following Boltzmann equation\cite{Abrikosov88,Mahan81},
\begin{eqnarray}\label{Boltzmann-equation-1}
{\partial {\bf k}\over\partial t}\cdot\nabla_{\bf k}f({\bf k})
=\left ({df\over dt}\right )_{\rm collisions}
\end{eqnarray}
where the right-hand side is the time rate of change due to the electron-electron collision,
while the factor ${\partial {\bf k}/\partial t}$ is equivalent to an acceleration, which is
equal to the forces on the electrons as,
\begin{eqnarray}\label{momentum-equation}
{\partial {\bf k}\over\partial t}=-e{\bf E},
\end{eqnarray}
with the charge $e$, where for a convenience in the following discussions, the magnetic
field has been dropped, i.e., ${\bf H}=0$, while only an electric field ${\bf E}$ is
applied to the system. In this case, we substitute Eq. (\ref{momentum-equation}) into
Eq. (\ref{Boltzmann-equation-1}), and rewrite the Boltzmann equation
(\ref{Boltzmann-equation-1}) as,
\begin{eqnarray}\label{Boltzmann-equation-2}
e{\bf E}\cdot\nabla_{\bf k}f({\bf k})+\left ({df\over dt}\right )_{\rm collisions}=0.
\end{eqnarray}
Following the discussions in Refs. \onlinecite{Prange64} and \onlinecite{Lee21}, we now
introduce the linear perturbation from the equilibrium in terms of the distribution function
as,
\begin{eqnarray}\label{distribution-function}
f({\bf k})&=&n_{\rm F}({\bar{\varepsilon}_{\bf k}})-{d n_{\rm F}({\bar{\varepsilon}_{\bf k}})
\over d{\bar{\varepsilon}_{\bf k}}}\tilde{\Phi}({\bf k}),
\end{eqnarray}
where $\tilde{\Phi}({\bf k})$ has been interpreted as a local shift of the chemical potential
at a given patch of EFS \cite{Lee21,Prange64}, and satisfies an antisymmetric relation
$\tilde{\Phi}(-{\bf k})=-\tilde{\Phi}({\bf k})$. With the help of the above distribution
function (\ref{distribution-function}), the Boltzmann equation (\ref{Boltzmann-equation-2})
can be linearized with the result that can be expressed explicitly as,
\begin{eqnarray}\label{Boltzmann-equation-3}
e{\bf v}_{\bf k}\cdot{\bf E}{\partial n_{\rm F}({\bar{\varepsilon}_{\bf k}})\over
\partial {\bar{\varepsilon}_{\bf k}}}=-\left ({df\over dt}\right )_{\rm collisions}
=I_{\rm e-e},
\end{eqnarray}
where ${\bf v}_{\bf k}=\nabla_{\bf k}{\bar{\varepsilon}_{\bf k}}$ is the electron velocity
and $I_{\rm e-e}$ is the electron-electron collision term.

\subsection{Electron umklapp scattering}\label{Umklapp-scattering}

\begin{figure}[h!]
\centering
\includegraphics[scale=0.65]{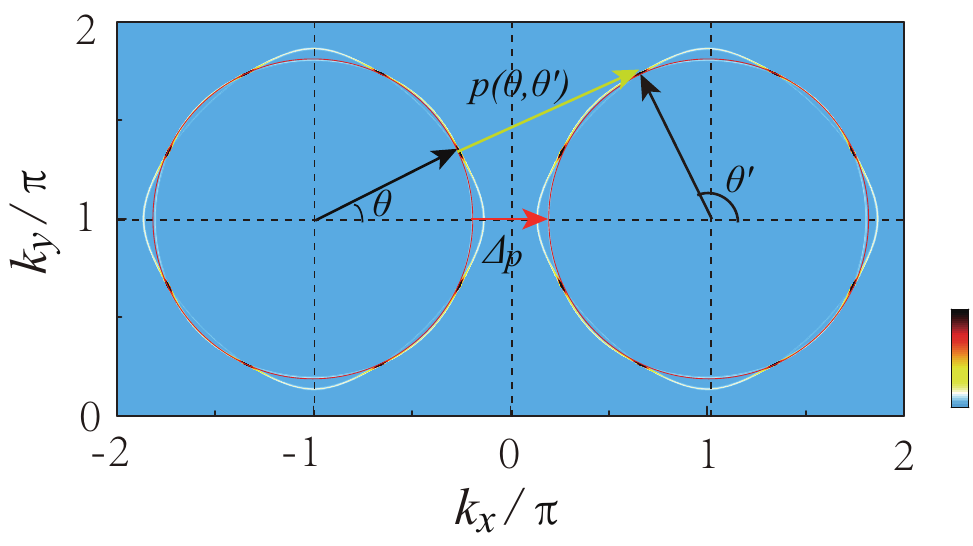}
\caption{(Color online) Illustration of the umklapp scattering process \cite{Lee21} in which
an electron on a circular electron Fermi surface (left) is scattered by its partner on the
umklapp electron Fermi surface (right), where the intensity map of the electron Fermi
surface is the same as shown in Fig. \ref{EFS-map}a, while the perfect circle (red) is the
circle with the radius ${\rm k}^{\rm TFA}_{\rm F}$, where ${\bf k}^{\rm TFA}_{\rm F}$ is the
Fermi wave vector of the tips of the Fermi arcs. An electron on the electron Fermi surface
(left) parametrized by the Fermi angle $\theta$ is scattered to a point parametrized by the
Fermi angle $\theta'$ on the umklapp electron Fermi surface (right) by the spin excitation
carrying momentum ${\rm p}(\theta,\theta')$. The minimal umklapp vector is $\Delta_{p}$ at
the antinode (the Fermi angle $\theta=0$). This physical picture is repeated for the other
three umklapp electron Fermi surfaces that are closest to the original electron Fermi
surface. \label{scatter-process}}
\end{figure}

\begin{figure}[h!]
\centering
\includegraphics[scale=0.5]{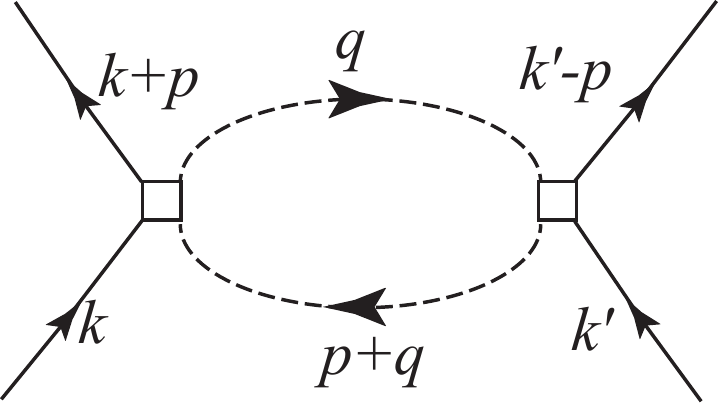}
\caption{The skeletal diagram of the umklapp scattering process for scattering electrons
from the spin excitation. The solid-line represents the electron propagator $G$, and the
dashed-line depicts the spin propagator $D^{(0)}$, while $\square$ describes the bare
vertex function $\Lambda$. \label{Feynman-diagram}}
\end{figure}

For evaluating the electron-electron collision in the Boltzmann equation
(\ref{Boltzmann-equation-3}), the mechanism of the momentum relaxation needs to be
introduced\cite{Abrikosov88,Mahan81}. After intensive investigations over more than three
decades, although the mechanism of the momentum relaxation for the T-linear resistivity
still remains controversial, the electron umklapp scattering is believed to be at the heart
of the striking behaviour of the electrical transport in the strange-metal phase of
overdoped cuprate superconductors
\cite{Hussey03,Rice17,Lee21,Honerkamp01,Hartnoll12,Tabis21}. In this paper, we adopt the
electron umklapp scattering as the mechanism of the momentum relaxation, and then study the
electrical transport in the strange-metal phase of overdoped cuprate superconductors. For a
convenience in the following discussions, the schematic picture for the electron umklapp
scattering process\cite{Lee21} is illustrated in Fig. \ref{scatter-process}, where an
electron on a circular EFS (left) is scattered by its partner on the umklapp EFS (right).
In Fig. \ref{scatter-process}, the intensity map of EFS is the same as in
Fig. \ref{EFS-map}a, while the perfect circle (red) is the circle with the radius
${\rm k}^{\rm TFA}_{\rm F}$, where ${\bf k}^{\rm TFA}_{\rm F}$ is the Fermi wave vector of
the tips of the Fermi arcs. This perfect circle EFS (red) connects all eight tips of the
Fermi arcs, and thus shows that almost all the spectral weight of the electron
excitation spectrum is accommodated on this circle EFS. It should be emphasized that in the
present case, the electron umklapp scattering is mediated by the same spin excitation as
in the case of the electron scattering for the renormalization of the electrons in
subsection \ref{Effective-propagator}. To understand the umklapp scattering mechanism in
the present case more clearly, the skeletal diagram of this umklapp scattering mechanism in
energy and momentum space is drawn in Fig. \ref{Feynman-diagram}.

In the recently pioneering work\cite{Lee21}, a model of the electrons scattered by a
critical bosonic mode with the umklapp scattering being the dominant momentum relaxation
mechanism has been studied, where the umklapp scattering process is described as an
electron-electron scattering via the exchange of the critical boson propagator, rather than
the scattering between electrons via the emission and absorption of bosons. Following their
discussions\cite{Lee21}, the electron-electron collision $I_{\rm e-e}$ in
Eq. (\ref{Boltzmann-equation-3}) originated from the electron umklapp scattering shown in
Fig. \ref{Feynman-diagram} can be evaluated as,
\begin{widetext}
\begin{eqnarray}\label{electron-collision-1}
I_{\rm e-e}&=&{1\over N^{2}}\sum_{{\bf k}',{\bf p}} {2\over T}
|P({\bf k},{\bf p},{\bf k}',\bar{\varepsilon}_{\bf k}
-\bar{\varepsilon}_{{\bf k}+{\bf p}+{\bf G}})|^{2}
\{\tilde{\Phi}({\bf k})+\tilde{\Phi}({\bf k'})
-\tilde{\Phi}({\bf k}+{\bf p}+{\bf G})-\tilde{\Phi}({\bf k}'-{\bf p})\}\nonumber\\
&\times&n_{\rm F}(\bar{\varepsilon}_{\bf k})n_{\rm F}(\bar{\varepsilon}_{{\bf k}'})
[1-n_{\rm F}(\bar{\varepsilon}_{{\bf k}+{\bf p}+{\bf G}})]
[1-n_{\rm F}(\bar{\varepsilon}_{{\bf k}'-{\bf p}})]
\delta(\bar{\varepsilon}_{\bf k}+\bar{\varepsilon}_{\bf k'}
-\bar{\varepsilon}_{{\bf k}+{\bf p}+{\bf G}}-\bar{\varepsilon}_{{\bf k}'-{\bf p}}),
\end{eqnarray}
\end{widetext}
which therefore leads to the appearance of the electrical resistance
\cite{Abrikosov88,Mahan81}, where ${\bf G}$ indicates a set of reciprocal lattice vectors,
and the umklapp scattering process in Eq. (\ref{electron-collision-1}) is described as an
electron-electron scattering via the exchange of the effective spin propagator
$P({\bf k},{\bf p},{\bf k}',\omega)$, rather than the scattering between electrons via the
emission and absorption of the spin excitation, while the effective spin propagator
$P({\bf k},{\bf p},{\bf k}',\omega)$ is obtained directly from Fig. \ref{Feynman-diagram}
as,
\begin{eqnarray}\label{ESP}
P({\bf k},{\bf p},{\bf k}',\omega)&=&{1\over N}\sum_{\bf q}\Lambda_{{\bf p}+{\bf q}+{\bf k}}
\Lambda_{{\bf q}+{\bf k}'}\Pi({\bf p},{\bf q},\omega).~~~
\end{eqnarray}
The reason of the electron-electron scattering via the exchange of the effective spin
propagator in the present case is the same as in the case discussed in
Ref. \onlinecite{Lee21}. For the normal scattering (${\bf G}=0$), the conservation of the
total momentum in Eq. (\ref{electron-collision-1}) is satisfied straightforwardly
\cite{Lee21}, since the distribution in the case of the normal scattering will rapidly
equilibrate to a fermion distribution function with a shifted overall momentum
$\tilde{\Phi}({\bf k})\propto {\bf k}\cdot{\bf E}$, which leads to that its contribution to
the integral of the electron-electron collision in Eq. (\ref{electron-collision-1}) is
exactly zero. However, if we consider the scattering between electrons via the emission and
absorption of the spin excitation, we would have to keep track of the extra shifted boson
distribution function as well, which introduces more complications \cite{Lee21}. Moreover,
it has been shown that the vanishing of the normal scattering in the electron-electron
collision (\ref{electron-collision-1}) is more general \cite{Lee21}. This is following a
basic fact that in order to stay on EFS and conserve the total momentum and energy, the
momentum of the normal scattering partner ${\bf k}'$ must equal to either ${\bf k}+{\bf p}$
or $-{\bf k}$. In the former case, the last two terms in $\{...\}$ in
Eq. (\ref{electron-collision-1}) cancel the first two terms. However, in the latter case,
since the antisymmetric relation $\tilde{\Phi}(-{\bf k})=-\tilde{\Phi}({\bf k})$, the first
two terms in $\{...\}$ in Eq. (\ref{electron-collision-1}) cancel, while the same
cancellation is valid for the last two terms corresponding to the outgoing pair
${\bf k}+{\bf p}$ and ${\bf k}'-{\bf p}$. These results therefore indicate that the
contribution from the normal scattering to the integral of the electron-electron collision
(\ref{electron-collision-1}) is negligible \cite{Lee21}.

In the usual case \cite{Abrikosov88,Mahan81}, the derivation of the Boltzmann equation
starting from the nonequilibrium electron propagator involves integrating over energy
$\omega$. However, the electrons at the bottom of the band (then the deep inside EFS) can
not be thermally excited, and as a matter of the principle, all the low-temperature
conduction processes in the strange-metals should involve only states at around EFS
\cite{Haldane18}. In particular, in the early discussions\cite{Prange64}, it has been
realized to pick a patch of EFS specified by $k(\theta)$ with the range $\theta\in [0,2\pi]$,
which defines a contour along EFS parametrized by the direction $\theta$ of the Fermi
momentum vector and integrate the perpendicular momentum and hence over
$\bar{\varepsilon}_{\bf k}$ instead. This is a formula expressed entirely in terms of the
EFS property. Furthermore, this method has been employed to study the low-temperature
T-linear resistivity due to the umklapp scattering from a critical bosonic mode\cite{Lee21}.
In the present case of the electron umklapp scattering mediated by the spin excitation, an
electron on EFS parametrized by the Fermi angle $\theta$ is scattered to a point
parametrized by the Fermi angle $\theta'$ on the umklapp EFS by the spin excitation carrying
momentum ${\rm p}(\theta,\theta')$ as shown in Fig. \ref{scatter-process}. In this case, the
usual distribution function $f({\bf k})$ can be replaced as $f[k(\theta)]$. However, in the
usual formulation, the vector ${\bf k}$ is decomposed into $k(\theta)$ and the momentum in
the perpendicular direction \cite{Lee21,Prange64}, which is then represented by
$\bar{\varepsilon}_{\bf k}$. From this EFS parametrization, the standard Boltzmann equation
(\ref{Boltzmann-equation-3}) now can be expressed simply, where the component of the
momentum ${\bf k}$ perpendicular to EFS is replaced by $\bar{\varepsilon}_{\bf k}/v_{\bf k}$
and $\bar{\varepsilon}_{\bf k}$ in turn is replaced by $\omega$. After a straightforward
calculation [see Appendix \ref{electron-electron-collision}], the above electron-electron
collision $I_{\rm e-e}$ in Eq. (\ref{electron-collision-1}) can be derived explicitly,
and then the Boltzmann equation (\ref{Boltzmann-equation-3}) can be obtained as,
\begin{eqnarray}\label{electron-collision}
e{\bf v}_{\rm F}(\theta)\cdot {\bf E}=-2\int {d\theta'\over {2\pi}}\zeta(\theta')
F(\theta,\theta')[\Phi(\theta)-\Phi(\theta')],~~~~~
\end{eqnarray}
where $\Phi(\theta)$ is defined as $\tilde{\Phi}[{\rm k}(\theta)]$ and satisfies an
antisymmetric relation \cite{Lee21} $\Phi(\theta)=-\Phi(\theta+\pi)$,
${\bf v}_{\rm F}(\theta)$ is the Fermi velocity at the Fermi angle $\theta$, and
$\zeta(\theta')={\rm k}^{2}_{\rm F}/[4\pi^{2}{\rm v}^{3}_{\rm F}]$ is the density of states
factor at angle $\theta'$ with the Fermi wave vector ${\rm k}_{\rm F}$ and Fermi velocity
${\rm v}_{\rm F}$, while the coefficient of $\Phi(\theta)$ in the first term of the
right-hand side of Eq. (\ref{electron-collision}),
\begin{eqnarray}\label{scattering-rate}
\gamma(\theta)=2\int {d \theta' \over {2\pi}} \zeta(\theta')F(\theta,\theta'),
\end{eqnarray}
is defined as the scattering out rate \cite{Lee21} from the state of ${\rm k}(\theta)$,
which also is so-called as {\it the angular dependence of the transport scattering rate},
while the kernel function $F(\theta,\theta')$ depends on the Fermi angles $\theta$ and
$\theta'$ in terms of the magnitude of the momentum transfer ${\rm p}(\theta,\theta')$,
i.e., $F(\theta,\theta')$ connects the points $\theta$ and $\theta'$ on the umklapp EFS as
shown in Fig. \ref{scatter-process}, and is given by,
\begin{eqnarray}\label{kernel-function}
F(\theta,\theta')&=&{1\over T}\int {d\omega\over 2\pi}{\omega^{2}
\over {\rm p}(\theta,\theta')}
{|\bar{P}[{\rm k}(\theta),{\rm p}(\theta,\theta'),\omega]|}^{2}\nonumber\\
&\times& n_{\rm B}(\omega)[1+n_{\rm B}(\omega)],~~~~~~
\end{eqnarray}
where the reduced effective spin propagator
$\bar{P}[{\rm k}(\theta),{\rm p}(\theta,\theta'),\omega]$ has been given in Appendix
\ref{electron-electron-collision}. This kernel function $F(\theta,\theta')$ can be also
called as the probability weight or the strength of the umklapp scattering.


\section{Low-temperature T-linear resistivity}\label{electron-resistivity}

The dc conductivity then is evaluated in a standard way by the momentum (then the Fermi
angle $\theta$) integral of the umklapp scattering process on EFS, where the current density
is given by\cite{Abrikosov88,Mahan81},
\begin{eqnarray}\label{current-density-1}
{\bf J}=-en_{0}{1\over N}\sum_{\bf k}{\bf v}_{\bf k}f({\bf k}),
\end{eqnarray}
with the momentum relaxation that is generated by the action of the electric field
on the mobile electrons \cite{Haldane18} at around EFS with the density $n_{0}$.
Substituting the distribution function $f({\bf k})$ in Eq. (\ref{distribution-function})
into the above current density equation (\ref{current-density-1}) and performing the radial
integration, the current density now can be obtained as,
\begin{eqnarray}\label{current-density}
{\bf J} &=& en_{0}{1\over N}\sum_{\bf k}{\bf v}_{\bf k}
{dn_{\rm F}({\bar{\varepsilon}_{\bf k}})\over d\bar{\varepsilon}_{\bf k}}
\tilde{\Phi}({\bf k})\nonumber\\
&=&-en_{0}{{\rm k}_{\rm F}\over {\rm v}_{\rm F}}\int {d\theta\over (2\pi)^{2}}
{\bf v}_{\rm F}(\theta)\Phi(\theta).~~~~~
\end{eqnarray}

For deriving the dc conductivity, we need to obtain the local shift of the
chemical potential $\Phi(\theta)$. The spectral weight of
${\rm Im}P({\bf k}_{\rm F},{\bf p}-{\bf k}_{\rm F},{\bf k'}_{\rm F},\omega)$ in
Eq. (\ref{electron-collision-1}) achieves its maximal value at around the antinodal region
[see Fig. \ref{effective-spin-propagator} in Appendix \ref{electron-electron-collision}],
where the scattering probability for two electrons is largest. In other words, the main
contribution to the kernel function $F(\theta,\theta')$ comes from
such umklapp scattering process in which the electron at around the antinodal region of the
circular EFS (left) shown in Fig. \ref{scatter-process} is scattered by its partner at
around the antinodal region of the umklapp circular EFS (right), where the Fermi angle
$\theta'$ is almost identical with the Fermi angle $\pi-\theta$, and then according to the
antisymmetric relation satisfied by $\Phi(\theta)$, the following relation,
\begin{eqnarray}\label{Phi-theta-1}
\Phi(\theta')=\Phi(\pi-\theta)=-\Phi(\theta),
\end{eqnarray}
is valid. In this relaxation-time approximation, the local shift of the chemical potential
$\Phi(\theta)$ can be derived straightforwardly from Eqs. (\ref{electron-collision})
and (\ref{scattering-rate}) as \cite{Lee21},
\begin{eqnarray}\label{Phi-theta}
\Phi(\theta)=-{e{\rm v}_{\rm F}{\rm cos}(\theta)E_{\hat{x}}\over 2\gamma(\theta)},
\end{eqnarray}
where the electric field ${\bf E}$ has been chosen along the $\hat{x}$-axis. Substituting
the above result of $\Phi(\theta)$ into Eq. (\ref{current-density}), the dc conductivity
therefore is obtained explicitly as,
\begin{eqnarray}\label{dc-conductivity}
\sigma_{\rm dc}(T)={1\over 2}e^{2}n_{0}{\rm k}_{\rm F}{\rm v}_{\rm F}
\int {d\theta\over (2\pi)^{2}}{\rm cos}^{2}(\theta)
{1\over\gamma(\theta)},
\end{eqnarray}
and then the resistivity is obtained directly from the above dc conductivity as,
\begin{eqnarray}\label{dc-resistivity}
\rho(T)={1\over \sigma_{\rm dc}(T)}.
\end{eqnarray}

\begin{figure}[h!]
\centering
\includegraphics[scale=0.95]{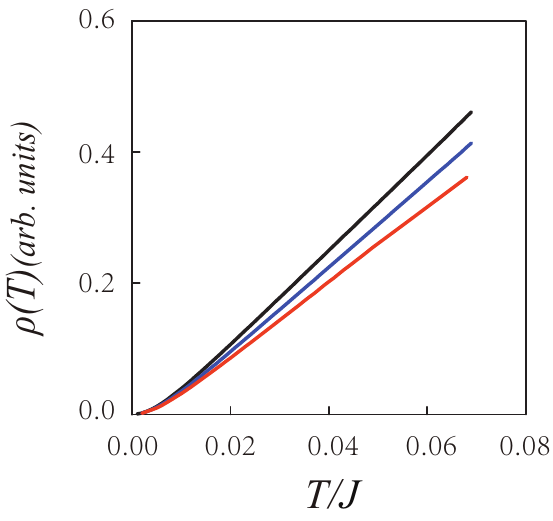}
\caption{(Color online) The resistivity as a function of temperature at $\delta=0.15$
(black-line), $\delta=0.18$ (blue-line), and $\delta=0.24$ (red-line).
\label{resistivity-temperature}}
\end{figure}
\begin{figure}[h!]
\centering
\includegraphics[scale=0.85]{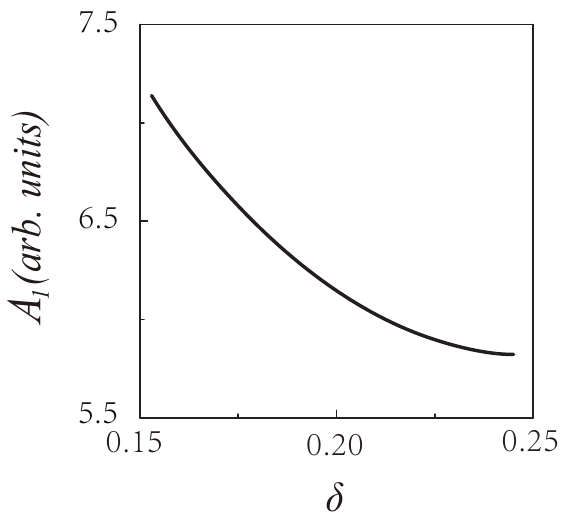}
\caption{T-linear resistivity coefficient as a function of doping.
\label{coefficient-doping}}
\end{figure}

Now we are ready to discuss the striking features of the electrical transport in the
strange-metal phase of overdoped cuprate superconductors. We have made a series of
calculations for the resistivity $\rho(T)$ in Eq. (\ref{dc-resistivity}) at different doping
levels, and the results of the resistivity $\rho(T)$ as a function of temperature at the
doping concentrations $\delta=0.15$ (black-line), $\delta=0.18$ (blue-line), and
$\delta=0.24$ (red-line) are plotted in Fig. \ref{resistivity-temperature}. Apparently, the
experimental results of the doping dependence of the low-temperature resistivity
\cite{Allen89,Gurvitch87,Takagi92,Martin90,Mandrus92,Ando01,Daou09,Cooper09,Legros19,Yuan22,Ayres21,Grisso21}
are qualitatively reproduced, where the highly unconventional features can be summarized as:
(i) the resistivity $\rho(T)$ as a function of temperature is a perfect straight line down
to the temperature $T\sim 0.015J=15$K; (ii) the low-temperature T-linear resistivity extends
over a wide doping range in the overdoped regime, where the T-linear resistivity coefficient
$A_{1}$ (then the strength of the T-linear resistivity) decreases with the increase of doping.
To see this doping dependence of the T-linear resistivity coefficient more clearly, we plot
$A_{1}$ versus doping $\delta$ in Fig. \ref{coefficient-doping}, where $A_{1}$
{\it increases} monotonically as the doping concentration is reduced in the overdoped regime,
this tendency of the doping dependence is in qualitative agreement with the experimental
observations on overdoped cuprate superconductors\cite{Legros19,Ayres21}; (iii) however, the
resistivity deviates from the pure T-linearity in the far-lower-temperature region
$T< 0.015J=15$K, while our numerical fit indicates that in this far-lower-temperature region,
the resistivity decreases quadratically as the temperature decreases. The results in
Fig. \ref{resistivity-temperature} therefore also indicate that the same spin excitation that
acts like a bosonic glue to hold the electron pairs together responsible for
superconductivity \cite{Feng15,Feng0306,Feng12,Feng15a} also dominates the electron umklapp
scattering responsible for the low-temperature T-linear resistivity in the strange-metal
phase.

Finally, it should be emphasized that the local shift of the chemical potential
$\Phi(\theta)$ can be also evaluated directly by the numerical solution of the Boltzmann
equation (\ref{electron-collision}) together with an additional electron-impurity collision
without making the relaxation-time approximation \cite{Lee21}, where the Fermi angle
$\theta'$ variable in Eq. (\ref{electron-collision}) can be discretized, and then the
integral-differential equation (\ref{electron-collision}) is converted into the matrix
equation. The accurate result of $\Phi(\theta)$ is obtained in terms of the numerical
calculation of the inverse of this matrix. In this case, we have also performed a numerical
calculation $\Phi(\theta)$ [then $\rho(T)$], and the results show that although the
resistivity saturates to a constant $\rho_{0}(T)$ induced by the impurity, the qualitative
behaviour of the resistivity is the same as that obtained in the above relaxation-time
approximation except for the subtle difference of slopes, which is also qualitatively
consistent with the results obtained from the electron umklapp scattering mediated by a
critical bosonic mode\cite{Lee21}.

\begin{figure}[h!]
\centering
\includegraphics[scale=0.85]{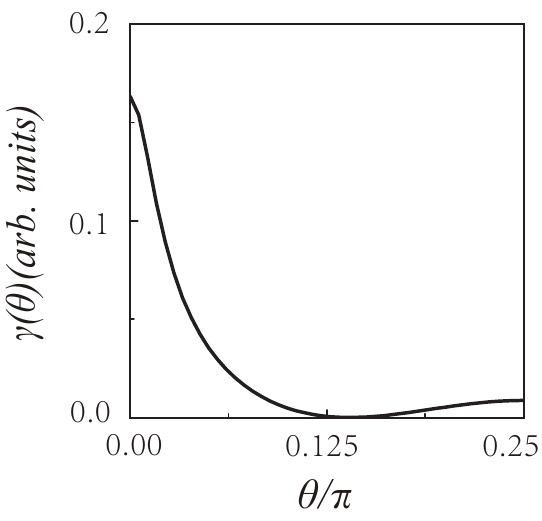}
\caption{The transport scattering rate $\gamma(\theta)$ as a function of Fermi angle
$\theta$ at $\delta=0.18$ with $T=0.05J$. \label{out-rate-theta}}
\end{figure}

From the dc conductivity in Eq. (\ref{dc-conductivity}) [then the resistivity in
Eq. (\ref{dc-resistivity})], it thus shows that the low-temperature T-linear resistivity in
the strange-metal phase can be attributed to the angle and temperature dependence of the
transport scattering rate $\gamma(\theta,T)$ in Eq. (\ref{scattering-rate}). To see this
point more clearly, we first plot $\gamma(\theta)$ as a function of Fermi angle $\theta$ at
$\delta=0.18$ with $T=0.05J$ in Fig. \ref{out-rate-theta}. In a comparison with the
corresponding angular dependence of the single-particle scattering rate $\Gamma(\theta)$ in
Fig. \ref{single-particle-scattering}, it shows that although the magnitude of
$\gamma(\theta)$ at an any given Fermi angle is less than that of $\Gamma(\theta)$ at the
corresponding Fermi angle, the global behaviour of the angular dependence of
$\gamma(\theta)$ is similar to that of $\Gamma(\theta)$, where $\gamma(\theta)$ is largest
at around the antinodal region, and smallest at around the tips of the Fermi arcs, which
is also consistent with the strong momentum dependence of the effective spin propagator
$P({\bf k},{\bf p}-{\bf k},{\bf k}',\omega)$ shown in Fig. \ref{effective-spin-propagator}
in Appendix \ref{electron-electron-collision}. In other words, both the transport scattering
rate $\gamma(\theta)$ and single-particle scattering rate $\Gamma(\theta)$ as a function of
Fermi angle presents a similar behavior of the effective spin propagator
$P({\bf k},{\bf p}-{\bf k},{\bf k}',\omega)$. The result in Fig. \ref{out-rate-theta}
therefore also shows that the resistivity is mainly dominated by the transport scattering
rate at both the antinodal and nodal regions.

\begin{figure}[h!]
\centering
\includegraphics[scale=0.85]{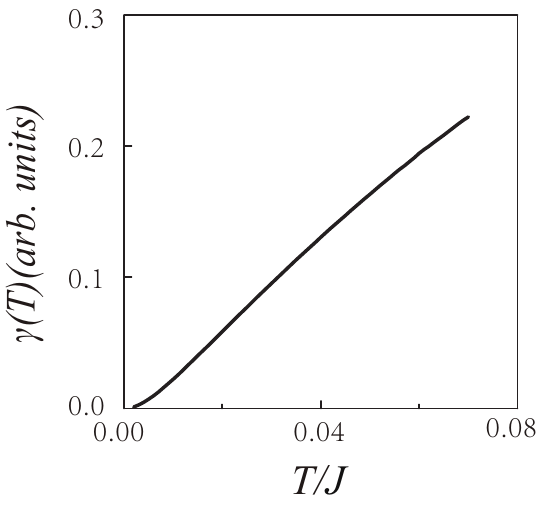}
\caption{The transport scattering rate $\gamma(T)$ at the antinode as a function of
temperature for $\delta=0.18$. \label{out-rate-temperature}}
\end{figure}

On the other hand, for an any given Fermi angle $\theta$, $\gamma(\theta,T)$ varies strongly
with temperature. To see this temperature dependence of $\gamma(\theta,T)$ more clearly, we
plot $\gamma(T)$ as a function of temperature for $\delta=0.18$ at the antinode in
Fig. \ref{out-rate-temperature}. It is surprising that $\gamma(T)$ is entirely T-linear in
the low-temperature region $T> 0.015J=15$K, where it decreases linearly with temperature as
the temperature decreases to $T\sim 0.015J=15$K, while this transport scattering rate
$\gamma(T)$ is instead T-quadratic in the far-lower-temperature region $T< 0.015J=15K$.
Moreover, although $\gamma(\theta,T)$ is highly anisotropic in momentum-space, this
low-temperature T-linear $\gamma(\theta,T)$ occurs at an any given Fermi angle $\theta$
(then for all directions), in agreement with the experimental observations\cite{Grisso21}.
In a comparison with the corresponding results of the temperature dependence of the
resistivity shown in Fig. \ref{resistivity-temperature}, we therefore find that the
T-linear behaviour of $\gamma(T)$ together with the temperature region and the T-quadratic
behaviour of $\gamma(T)$ together with the temperature region are respectively the same as
the corresponding behaviours and regions in the resistivity $\rho(T)$, which shows clearly
that the T-linear resistivity with the temperature region and the T-quadratic resistivity
with the temperature region are governed respectively by the T-linear transport scattering
rate with the temperature region and T-quadratic transport scattering rate with the
temperature region.

\begin{figure}
\centering
\includegraphics[scale=1.0]{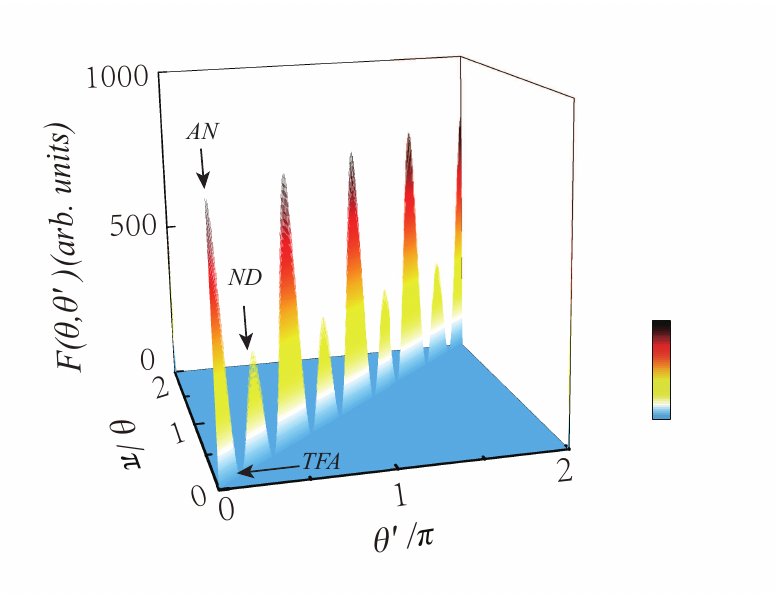}
\caption{(Color online) The surface plot of the kernel function $F(\theta,\theta')$ at
$\delta=0.18$ with $T=0.05J$, where AN, TFA, and ND denote the antinode, tip of the Fermi
arc, and node, respectively. \label{kernel}}
\end{figure}

For a further understanding of the nature of the transport scattering rate
$\gamma(\theta,T)$, we discuss the temperature dependence of the kernel function
$F(\theta,\theta')$ in Eq. (\ref{kernel-function}), since the temperature dependence of
$\gamma(\theta,T)$ in Eq. (\ref{scattering-rate}) is mainly determined by the temperature
dependence of $F(\theta,\theta')$. In Fig. \ref{kernel}, we plot the surface plot of
$F(\theta,\theta')$ at $\delta=0.18$ with $T=0.05J$, where the probability weight of the
electron umklapp scattering has been separated clearly into three characteristic regions:
(i) the antinodal region, where a particularly large fraction of the probability weight
is located, leading to that $\gamma(\theta)$ is largest at around the antinodal region;
(ii) the nodal region, where a small amount of the probability weight is inhabited,
leading to that the magnitude of $\gamma(\theta)$ at around the nodal region is much
smaller than that at around the antinodal region;
(iii) the region at around the tips of the Fermi arcs, where the
strength of the umklapp scattering is anomalously small, leading to the appearance of the
weakest scattering at around the tips of the Fermi arcs. The characteristic feature of the
tips of the Fermi arcs is that both the real and imaginary parts of the electron normal
self-energy have the anomalously small values\cite{Feng16}, indicating that the interaction
(then the scattering) between electrons at around the tips of the Fermi arcs is particularly
weak. In other words, although the electron density of states is largest at around the tips
of the Fermi arcs, the electron scattering at around the tips of the Fermi arcs is quite
weak, and then the electrons at around the tips of the Fermi arcs move more freely than
those at other parts of EFS. The above result in Fig. \ref{kernel} indicates that the
electron umklapp scattering is concentrated at around the antinodes and nodes, and therefore
is well consistent with the result of the angular dependence of $\gamma(\theta,T)$ shown in
Fig. \ref{out-rate-theta}. However, the strengths of the antinodal and nodal umklapp
scattering are temperature dependent, which induces a competition between the antinodal and
nodal umklapp scattering, and can be well understood in terms of the ratio of the strength
of the nodal umklapp scattering to the strength of the antinodal umklapp scattering,
\begin{eqnarray*}
R_{\rm F}(T)={F(\theta_{\rm ND},\theta'_{\rm ND},T)
\over F(\theta_{\rm AN},\theta'_{\rm AN},T)}.
\end{eqnarray*}
\begin{figure}[h!]
\centering
\includegraphics[scale=0.85]{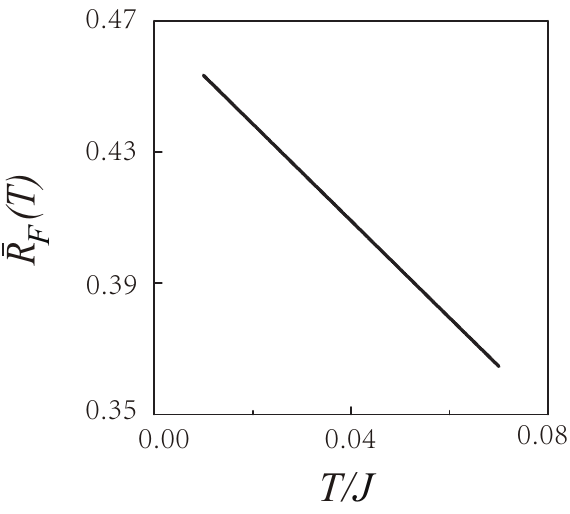}
\caption{The ratio of the strength of the nodal umklapp scattering to the strength of the
antinodal umklapp scattering as a function of temperature for $\delta=0.18$. \label{R-ratio}}
\end{figure}
However, as we have mentioned in subsection \ref{Effective-propagator}, the calculation in
this paper is performed numerically for a finite lattice, which leads to that the weight
of the $\delta$-function type peak in $F(\theta,\theta')$ at the antinode (node) spreads
on the extremely small area $\{\theta_{\rm AN}\}$ $[\{\theta_{\rm ND}\}]$ around the
antinode (node) as shown in Fig. \ref{kernel}. In particular, the summation of these spread
weights around this extremely small area $\{\theta_{\rm AN}\}$ $[\{\theta_{\rm ND}\}]$ is
less affected by the calculation for a finite lattice. In this case, a more appropriate
ratio can be obtained as,
\begin{eqnarray}\label{order-parameter}
\bar{R}_{\rm F}(T)={\bar{F}_{\rm ND}(T)\over \bar{F}_{\rm AN}(T)},
\end{eqnarray}
for the reduction of the size effect in the finite-lattice calculation, where
$\bar{F}_{\rm AN}(T)$ and $\bar{F}_{\rm ND}(T)$ are given by,
\begin{eqnarray*}
\bar{F}_{\rm AN}(T)&=&{1\over 2\pi}\sum_{\substack{\theta_{\rm AN}\in\{\theta_{\rm AN}\}\\
\theta'_{\rm AN}\in\{\theta'_{\rm AN}\}}}F(\theta,\theta',T),\\
\bar{F}_{\rm ND}(T)&=&{1\over 2\pi}\sum_{\substack{\theta_{\rm ND}\in\{\theta_{\rm ND}\}\\
\theta'_{\rm ND}\in\{\theta'_{\rm ND}\}}}F(\theta,\theta',T),
\end{eqnarray*}
with the summation $\theta_{\rm AN}\in\{\theta_{\rm AN}\}$
$[\theta'_{\rm AN}\in\{\theta'_{\rm AN}\}]$ that is restricted to the extremely small area
$\{\theta_{\rm AN}\}$ $[\{\theta_{\rm ND}\}]$ around the antinode (node). In this case, we
have carried out a series of calculations for the ratio $\bar{R}_{\rm F}(T)$ at different
doping levels, and the result of $\bar{R}_{\rm F}(T)$ as a function of temperature at
$\delta=0.18$ is plotted in Fig. \ref{R-ratio}, where $\bar{R}_{\rm F}(T)$ decreases
monotonically with the increase of temperature. In other words, although both the strengths
of the nodal and antinodal umklapp scattering decrease with the decrease of temperature,
the decrease of the strength of the nodal umklapp scattering is slower than that of the
antinodal umklapp scattering.

We now turn to show why the transport scattering rate $\gamma(T)$ [then the resistivity
$\rho(T)$] exhibits a crossover from the T-linear behaviour in the low-temperature region
into T-quadratic behaviour in the far-lower-temperature region? The expression form of
the kernel function $F(\theta,\theta')$ in Eq. (\ref{kernel-function}) indicates that
$F(\theta,\theta')$ is proportional to the effective spin propagator
$P({\bf k},{\bf p},{\bf k}',\omega)$. For a convenience in the following discussions, the
effective spin propagator $P({\bf k},{\bf p},{\bf k}',\omega)$ in Eq. (\ref{ESP}) can be
rewritten as,
\begin{eqnarray}\label{reduced-propagator}
P({\bf k},{\bf p},{\bf k}',\omega)=-{1\over N}\sum\limits_{{\bf q},j}(-1)^{j+1}
{\varpi_{j}({\bf k},{\bf p},{\bf k}',{\bf q})\over\omega^{2}
-[\omega^{(j)}_{{\bf p}{\bf q}}]^{2}},
~~~~~
\end{eqnarray}
with $j=1,2$, the weight functions $\varpi_{1}({\bf k},{\bf p},{\bf k}',{\bf q})$ and
$\varpi_{2}({\bf k},{\bf p},{\bf k}',{\bf q})$ that are given by,
\begin{subequations}
\begin{eqnarray}
\varpi_{1}({\bf k},{\bf p},{\bf k}',{\bf q})&=& \Lambda_{{\bf k}+{\bf p}+{\bf q}}
\Lambda_{{\bf q}+{\bf k}'}\bar{W}^{(1)}_{{\bf p}{\bf q}},\\
\varpi_{2}({\bf k},{\bf p},{\bf k}',{\bf q})&=& \Lambda_{{\bf k}+{\bf p}+{\bf q}}
\Lambda_{{\bf q}+{\bf k}'}\bar{W}^{(2)}_{{\bf p}{\bf q}}.
\end{eqnarray}
\end{subequations}
Substituting above result in Eq. (\ref{reduced-propagator}) into Eq. (\ref{kernel-function}),
the kernel function $F(\theta,\theta')$ can be rewritten as,
\begin{widetext}
\begin{eqnarray}\label{reduced-kernel-1}
F(\theta,\theta') &=& {1\over T}{1\over N^{2}}\sum\limits_{{\bf q},{\bf q}',jj'}(-1)^{j+j'}
P\int_{-\infty}^{\infty}{d\omega\over 2\pi}{\omega^{2}\over p(\theta,\theta')}
{e^{\beta\omega}\over [e^{\beta\omega}-1]^{2}}{\varpi_{j}(\theta,\theta',{\bf q})
\over\omega^{2}-[\omega^{(j)}_{\theta\theta'}({\bf q})]^{2}}
{\varpi_{j'}(\theta,\theta',{\bf q}')\over
\omega^{2}-[\omega^{(j')}_{\theta\theta'}({\bf q}')]^{2}}\nonumber
\end{eqnarray}
\begin{eqnarray}
&=& {1\over N^{2}}\sum\limits_{{\bf q},{\bf q}',jj'}(-1)^{j+j'}
{\varpi_{j}(\theta,\theta',{\bf q})\varpi_{j'}(\theta,\theta',{\bf q}')
\over 2\pi p(\theta,\theta')}{1\over T^{2}}
i2\pi I[C_{j}(\theta,\theta',{\bf q}),C_{j'}(\theta,\theta',{\bf q}')],~~~~~
\end{eqnarray}
with $\varpi_{j}(\theta,\theta',{\bf q})=\varpi_{j}
[k(\theta),p(\theta,\theta'),{\bf k}_{\rm F}',{\bf q}]$, $\varpi_{j'}(\theta,\theta',{\bf q})
=\varpi_{j'}[k(\theta),p(\theta,\theta'),{\bf k}_{\rm F}',{\bf q}']$,
$\omega^{(j)}_{\theta\theta'}({\bf q})=\omega^{(j)}_{p(\theta,\theta'){\bf q}}$,
$\omega^{(j')}_{\theta\theta'}({\bf q}')=\omega^{(j')}_{p(\theta,\theta'){\bf q}'}$,
and the function,
\begin{eqnarray}\label{C-function-1}
I_{jj'}(\theta,\theta',{\bf q},{\bf q}')=P\int_{-\infty}^{\infty}{dx\over i2\pi}
{e^{x}\over (e^{x}-1)^{2}}{x^{2}\over [x^{2}-C^{2}_{j}(\theta,\theta',{\bf q})]
[x^{2}-C^{2}_{j'}(\theta,\theta',{\bf q}')]},~~~~
\end{eqnarray}
\end{widetext}
where $C_{j}(\theta,\theta',{\bf q})$ and $C_{j'}(\theta,\theta',{\bf q}')$ are defined as
$C_{j}(\theta,\theta',{\bf q})=\omega^{(j)}_{\theta\theta'}({\bf q})/T$ and
$C_{j'}(\theta,\theta',{\bf q}')=\omega^{(j')}_{\theta\theta'}({\bf q}')/T$, respectively.
From the MF spin propagator in Eq. (\ref{SGF-1}), the spin spectral function can be
obtained directly as,
\begin{eqnarray}\label{spin-spectral-function}
A_{\rm spin}({\bf k})=\pi {B_{\bf k}\over\omega_{\bf k}}[\delta(\omega-\omega_{\bf k})
-\delta(\omega+\omega_{\bf k})],
\end{eqnarray}
where the MF spin excitation energy spectrum $\omega_{\bf k}$ is an even function
of momentum ${\bf k}$, and in particular, $\omega_{\bf k}|_{{\bf k}={\bf k}_{\rm A}}$ has an
extremely small value \cite{Feng15} [$\omega_{{\bf k}_{\rm A}}=0.00205J\approx 2$K at doping
$\delta=0.18$], where ${\bf k}_{\rm A}=[\pm\pi,\pm\pi]$ is the antiferromagnetic wave vector,
which therefore leads to that the spin excitations at around the ${\bf k}_{\rm A}$ point of
BZ have the largest density of states, and then the spin response is mainly governed by
these spin excitations. In this case, the effective spin excitation energy dispersions
$\omega^{(1)}_{{\bf p}{\bf q}}$ and $\omega^{(2)}_{{\bf p}{\bf q}}$ in
Eq. (\ref{spin-bubble}) can be expanded, and then can be expressed approximately as,
\begin{subequations}\label{effective-spin-excitation}
\begin{eqnarray}
\omega^{(1)}_{{\bf p}{\bf q}}&=&\omega_{{\bf q}+{\bf p}}+\omega_{\bf q}
\approx a({\bf q})p^{2}+2\omega_{\bf q},\\
\omega^{(2)}_{{\bf p}{\bf q}}&=&\omega_{{\bf q}+{\bf p}}-\omega_{\bf q}
\approx a({\bf q})p^{2},
\end{eqnarray}
\end{subequations}
where $a({\bf q})=(d^{2}\omega_{\bf q}/d^{2}{\bf q})$. The results in
Eq. (\ref{effective-spin-excitation}) therefore indicate that the effective spin propagator
$P({\bf k},{\bf p},{\bf k}',\omega)$ in Eq. (\ref{reduced-propagator}) is scaled by $p^{2}$.
However, due to this $p^{2}$ scaling in the effective spin propagator
(\ref{reduced-propagator}), the temperature scale when the electron umklapp scattering kicks
in can be very low \cite{Lee21}, being proportional to $\Delta^{2}_{p}$, where $\Delta_{p}$
is the minimal umklapp vector at the antinode shown in Fig. \ref{scatter-process}. In other
words, $\bar{a}\Delta^{2}_{p}$ is referred to as the temperature scale, i.e.,
$T_{\rm scale}=\bar{a}\Delta^{2}_{p}$, where $\bar{a}=(1/N)\sum\limits_{\bf q}a({\bf q})$
is the average value, and is a constant at a given doping. Moreover, this temperature scale
is obtained numerically as $T_{\rm scale}=\bar{a}\Delta^{2}_{p}=0.01513J\approx 15$K at
doping $\delta=0.18$, which is well consistent with the crossover temperature shown in
Fig. \ref{resistivity-temperature}. In this case, the function
$I_{jj'}(\theta,\theta',{\bf q},{\bf q}')$ in the above equation (\ref{C-function-1}) can
be derived respectively in three different temperature regions:\\
(i) in the region for finite values of $C_{j}(\theta,\theta',{\bf q})$ and
$C_{j'}(\theta,\theta',{\bf q}')$, the function $I_{jj'}(\theta,\theta',{\bf q},{\bf q}')$
in Eq. (\ref{C-function-1}) can be evaluated [see Appendix \ref{reduced-kernel-function}]
as,
\begin{widetext}
\begin{equation}\label{C-function-2}
I_{jj'}(\theta,\theta',{\bf q},{\bf q}') = i{1\over 2[C^{2}_{j}(\theta,\theta',{\bf q})
-C^{2}_{j'}(\theta,\theta',{\bf q}')]}\left ( {C_{j}(\theta,\theta',{\bf q})
e^{C_{j}(\theta,\theta',{\bf q})}\over [e^{C_{j}(\theta,\theta',{\bf q})}-1]^{2}}
-{C_{j'}(\theta,\theta',{\bf q}')e^{C_{j'}(\theta,\theta',{\bf q}')}\over
[e^{C_{j'}(\theta,\theta',{\bf q}')}-1]^{2}}\right ).
\end{equation}
In particular, in the low-temperature region, which is corresponding to the case of
$T>[T_{\rm scale}+\omega_{{\bf k}_{\rm A}}]$, the function
$I_{jj'}(\theta,\theta',{\bf q},{\bf q}')$ in the above equation (\ref{C-function-2}) can
be reduced as
$I_{jj'}(\theta,\theta',{\bf q},{\bf q}')\sim -iT^{3}/[4(\bar{a}\Delta^{2}_{p})^{3}]$,
and then kernel function in Eq. (\ref{reduced-kernel-1}) in the low-temperature region is
obtained explicitly as,
\begin{eqnarray}\label{reduced-kernel-3}
F(\theta,\theta') &\approx& {1\over N^{2}}\sum\limits_{{\bf q},{\bf q}',jj'}(-1)^{j+j'}
{\varpi_{j}(\theta,\theta',{\bf q})\varpi_{j'}(\theta,\theta',{\bf q}')\over
2p(\theta,\theta')}
{T\over \omega^{(j)}_{\theta\theta'}({\bf q})\omega^{(j')}_{\theta\theta'}({\bf q}')
[\omega^{(j)}_{\theta\theta'}({\bf q})+\omega^{(j')}_{\theta\theta'}({\bf q}')]}\propto T,
~~~~~
\end{eqnarray}
which leads to the transport scattering rate $\gamma(T)\propto T$ [then the resistivity
$\rho(T)\propto T$] in the low-temperature region; \\
(ii) on the other hand, in the far-lower-temperature region, which is corresponding to
the case of $T<T_{\rm scale}$, the function $I_{jj'}(\theta,\theta',{\bf q},{\bf q}')$ in
Eq. (\ref{C-function-1}) can be derived straightforwardly as,
\begin{eqnarray}\label{C-function-4}
I_{jj'}(\theta,\theta',{\bf q},{\bf q}') &= &
-i{1\over 2\pi C^{2}_{j}(\theta,\theta',{\bf q})C^{2}_{j'}(\theta,\theta',{\bf q}')}
\int_{0}^{\infty}dx {e^{x}\over (e^{x}-1)^{2}}{x^{2}\over
[1-x^{2}/C^{2}_{j}(\theta,\theta',{\bf q})][1-x^{2}/C^{2}_{j'}(\theta,\theta',{\bf q}')]}
\nonumber\\
&=&-i{1\over 2\pi C^{2}_{j}(\theta,\theta',{\bf q})C^{2}_{j'}(\theta,\theta',{\bf q}')}
\int_{0}^{\infty}dx{e^{x}x^{2}\over (e^{x}-1)^{2}} \left [ 1+{x^{2}\over
C^{2}_{j}(\theta,\theta',{\bf q})}+\left ({x^{2}\over C^{2}_{j}(\theta,\theta',{\bf q})}
\right )^{2}+\cdots\right ]\nonumber
\end{eqnarray}
\begin{eqnarray}
&\times& \left [ 1+{x^{2}\over C^{2}_{j'}(\theta,\theta',{\bf q}')}+\left ({x^{2}\over
C^{2}_{j'}(\theta,\theta',{\bf q}')}\right )^{2}+\cdots\right ]
\approx -i{1\over 2\pi
C^{2}_{j}(\theta,\theta',{\bf q})C^{2}_{j'}(\theta,\theta',{\bf q}')}
\int_{0}^{\infty}dx{e^{x}x^{2}\over (e^{x}-1)^{2}}\nonumber\\
&=& -i{1\over 6C^{2}_{j}(\theta,\theta',{\bf q})C^{2}_{j'}(\theta,\theta',{\bf q}')}
=-i{T^{4}\over 6[\bar{a}\Delta^{2}_{p}]^{2}[\bar{a}\Delta^{2}_{p}
+\omega_{{\bf k}_{\rm A}}]^{2}},~~~~
\end{eqnarray}
and then the kernel function in Eq. (\ref{reduced-kernel-1}) can be obtained in the
far-lower-temperature region as,
\begin{eqnarray}
F(\theta,\theta') \approx {1\over N^{2}}\sum\limits_{{\bf q},{\bf q}',jj'}(-1)^{j+j'}
{\varpi_{j}(\theta,\theta',{\bf q})\varpi_{j'}(\theta,\theta',{\bf q}')\over
6p(\theta,\theta')}{1\over T^{2}}{T^{4}\over [\bar{a}\Delta^{2}_{p}]^{2}
[\bar{a}\Delta^{2}_{p}+\omega_{{\bf k}_{\rm A}}]^{2}}\propto T^{2},~~~~
\end{eqnarray}
\end{widetext}
which naturally gives rise to a T-quadratic behaviour of the transport scattering rate
$\gamma(T)\propto T^{2}$ [then the resistivity $\rho(T)\propto T^{2}$] in the
far-lower-temperature region;\\
(iii) however, in the temperature region of
$T_{\rm scale}<T<T_{\rm scale}+\omega_{{\bf k}_{\rm A}}$, which is corresponding to the
crossover region of the transport scattering rate. In this extremely narrow crossover region
(from $\sim$15K to $\sim$ 17K), the resistivity is neither T-linear nor T-quadratic, but is
a nonlinear in temperature. The above results therefore explain why the resistivity has
a crossover from the T-linear behaviour in the low-temperature region into the T-quadratic
behaviour in the far-lower-temperature region. The above results also show that the effect
of the umklapp scattering via the exchange of the effective spin propagator is not
exponentially small at the low-temperature region as in the case of the electron-phonon
coupling, but is power law down to the far-lower-temperature region as in the case of the
coupling of the electrons with a critical bosonic mode \cite{Lee21}.

\section{Summary and discussion}\label{summary}

Starting from the low-energy electronic structure of the strange-metal phase in overdoped
cuprate superconductors, we have studied the nature of the low-temperature electrical
transport, where the angular dependence of the transport scattering rate is arisen from
the umklapp scattering between electrons by the exchange of the effective spin propagator,
and is employed to evaluate the resistivity by making use of the Boltzmann
equation. Our results show that although the magnitude of the transport scattering rate at
an any given Fermi angle is smaller than the corresponding value of the single-particle
scattering rate, the transport scattering rate presents a similar behavior of the
single-particle scattering rate, where the transport scattering rate is largest at around
the antinodal region and smallest at around the tips of the Fermi arcs, indicating that
the resistivity is mainly dominated by the antinodal and nodal umklapp scattering. In
particular, a very low temperature $T_{\rm scale}$ scales with $\Delta^{2}_{p}$, this
leads to that in the low-temperature region ($T>T_{\rm scale}$), the transport scattering
rate is T-linear with the T-linear resistivity coefficient that decreases with the increase
of doping. However, in the far-lower-temperature region ($T<T_{\rm scale}$), the transport
scattering rate is instead T-quadratic. This T-linear behaviour of the transport scattering
rate in the low-temperature region and T-quadratic behaviour in the far-lower-temperature
region in turn induces respectively the T-linear resistivity in the low-temperature region
and T-quadratic resistivity in the far-lower-temperature region. Our theory also shows that
the same spin excitation that acts like a bosonic glue to hold the electron pairs together
responsible for the exceptionally high $T_{\rm c}$ also mediates the electron umklapp
scattering in the strange-metal phase of overdoped cuprate superconductors responsible for
the T-linear resistivity.

It should be emphasized that the effective spin propagator
$P({\bf k},{\bf p},{\bf k}',\omega)$ in Eq. (\ref{ESP}) is obtained in the MF level
\cite{Feng16,Feng15}, i.e., $P({\bf k},{\bf p},{\bf k}',\omega)$ in Eq. (\ref{ESP}) is
obtained as a convolution of two MF spin propagators in Eq. (\ref{SGF-1}). Thus the
umklapp scattering between electrons in Eq. (\ref{electron-collision-1}) by the exchange
of this effective MF spin propagator is better suited for the discussions of the electrical
transport in the strange-metal phase of overdoped cuprate superconductors as in case of
the umklapp scattering between electron by the exchange of the phenomenologically critical
bosonic mode\cite{Lee21}. This follows a basic fact that in the overdoped regime, the
effect of the magnetic fluctuation becomes quite weak
\cite{Fujita12,Timusk99,Kastner98,Eschrig06}, and then the scattering between electrons by
the exchange of the effective MF spin propagator can give a suitable description of the
renormalization of the electrons\cite{Feng16,Liu21,Cao21,Zeng22} and the related electrical
transport discussed in this paper. However, in the underdoped regime, where the magnetic
fluctuation is particularly strong\cite{Fujita12,Timusk99,Kastner98,Eschrig06} and can be
described in terms of the full spin propagator\cite{Kuang15,Yuan01,Feng98}, where the spin
self-energy is derived in terms of the charge-carrier bubble, and then the umklapp
scattering between electrons should be mediated by the exchange of the effective full spin
propagator for a suitable discussion of the electrical transport in the underdoped cuprate
superconductors. These and the related issues are under investigation now. The transport
scattering mechanism developed in this paper for the understanding of the electrical
transport in the strange-metal phase of overdoped cuprate superconductors can be also
employed to study the electrical transport in other families of strange metals
\cite{Bruin13,Grigera01} in the doped regime, where the magnetic fluctuation is rather
weak. In particular, based on this transport scattering theory, we have also discussed the
low-temperature T-linear resistivity of the electron-doped cuprate superconductors in the
overdoped regime\cite{Ma22}, where we show the common mechanism linking the electrical
transport of both the hole- and electron-doped cuprate superconductors in the overdoped
regime. These and the related works will be presented elsewhere.

\section*{Acknowledgements}

The authors would like to thank Dr. Lin Zhao and Dr. Xiang Li for the helpful discussions.
This work is supported by the National Key Research and Development Program of China under
Grant No. 2021YFA1401803, and the National Natural Science Foundation of China under Grant
Nos. 11974051, 12274036, and 12247116.


\begin{appendix}

\section{Derivation of electron-electron collision}\label{electron-electron-collision}

The aim of this Appendix is to derive the electron-electron collision $I_{\rm e-e}$ in
Eq. (\ref{electron-collision}) of the main text. The electron-electron collision in
Eq. (\ref{electron-collision-1}) can be also rewritten as,
\begin{widetext}
\begin{eqnarray}\label{electron-collision-2}
I_{\rm e-e}&=&{1\over N^{2}}\sum_{{\bf k}',{\bf p}} {2\over T}|P({\bf k},{\bf p}
-{\bf k},{\bf k}',\bar{\varepsilon}_{\bf k}-\bar{\varepsilon}_{{\bf p}+{\bf G}})|^{2}
[\tilde{\Phi}({\bf k})+\tilde{\Phi}({\bf k'})-\tilde{\Phi}({\bf p}+{\bf G})
-\tilde{\Phi}({\bf k}'+{\bf k}-{\bf p})] \nonumber\\
&\times& n_{\rm F}(\bar{\varepsilon}_{\bf k})n_{\rm F}(\bar{\varepsilon}_{{\bf k}'})
[1-n_{\rm F}(\bar{\varepsilon}_{{\bf p}+{\bf G}})]
[1-n_{\rm F}(\bar{\varepsilon}_{{\bf k}'+{\bf k}-{\bf p}})]
\delta(\bar{\varepsilon}_{\bf k}+\bar{\varepsilon}_{\bf k'}
-\bar{\varepsilon}_{{\bf p}+{\bf G}}-\bar{\varepsilon}_{{\bf k}'+{\bf k}-{\bf p}}),~~~~
\end{eqnarray}
with the effective spin propagator $P({\bf k},{\bf p}-{\bf k},{\bf k}',\omega)$ that can
be expressed explicitly as,
\begin{eqnarray}\label{ESP-2}
P({\bf k},{\bf p}-{\bf k},{\bf k}',\omega)={1\over N}\sum_{\bf q}\Lambda_{{\bf p}+{\bf q}}
\Lambda_{{\bf q}+{\bf k}'}\Pi({\bf p}-{\bf k},{\bf q},\omega)
=\int^{\infty}_{-\infty}{d\omega'\over\pi}
{{\rm Im}P({\bf k},{\bf p}-{\bf k},{\bf k}',\omega')\over\omega'-\omega}, ~~~~~~
\end{eqnarray}
\end{widetext}
where the imaginary part of the effective spin propagator
${\rm Im}P({\bf k},{\bf p}-{\bf k},{\bf k}',\omega)$ is directly related to the effective
spin spectral function, and is also defined as the scattering probability for two electrons.
\begin{figure}[h!]
\centering
\includegraphics[scale=0.80]{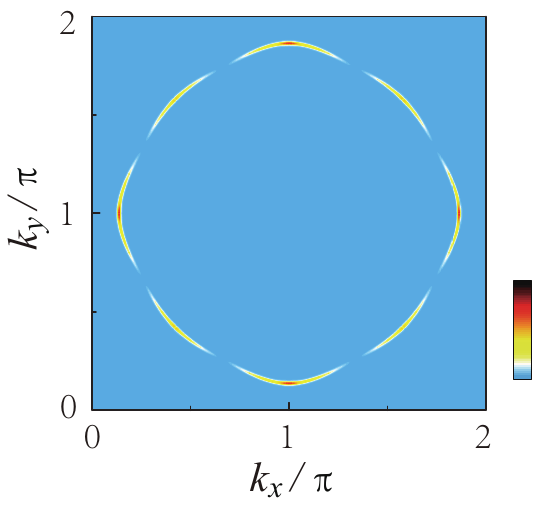}
\caption{(Color online) The intensity map of the imaginary part of the effective spin
propagator ${\rm Im}P({\bf k},{\bf p}-{\bf k},{\bf k}',\omega)$ along
${\bf k}={\bf k}'={\bf k}_{\rm F}$ in a $[p_{x},p_{y}]$ plane at $\delta=0.18$ for energy
$\omega=-0.05J$ with $T=0.002J$.\label{effective-spin-propagator}}
\end{figure}
However, in our previous discussions \cite{Mou19}, we have shown that for given momentums
${\bf k}$ and ${\bf k}'$, ${\rm Im}P({\bf k},{\bf p}-{\bf k},{\bf k}',\omega)$ exhibits a
remarkable evolution with momentum ${\bf p}$ and $\omega$ except for $\omega=0$, where
${\rm Im}P({\bf k},{\bf p}-{\bf k},{\bf k}',0)=0$. To see this unusual momentum ${\bf p}$
dependence of ${\rm Im}P({\bf k},{\bf p}-{\bf k},{\bf k}',\omega)$ more clearly, we plot
the intensity map of ${\rm Im}P({\bf k},{\bf p}-{\bf k},{\bf k}',\omega)$ along EFS
${\bf k}={\bf k}'={\bf k}_{\rm F}$ in a $[p_{x},p_{y}]$ plane at $\delta=0.18$ for energy
$\omega=-0.05J$ with $T=0.002J$ in Fig. \ref{effective-spin-propagator}, where the spectral
weight of ${\rm Im}P({\bf k}_{\rm F},{\bf p}-{\bf k}_{\rm F},{\bf k'}_{\rm F},\omega)$
along EFS ${\bf k}={\bf k}'={\bf k}_{\rm F}$ converges on the corresponding EFS
${\bf p}={\bf k}_{\bf F}$, i.e.,
${\rm Im}P({\bf k}_{\rm F},{\bf p}-{\bf k}_{\rm F},{\bf k'}_{\rm F},\omega)\neq 0$ for
${\bf p}={\bf k}_{\bf F} $, and otherwise
${\rm Im}P({\bf k}_{\rm F},{\bf p}-{\bf k}_{\rm F},{\bf k'}_{\rm F},\omega)=0$. In
particular, the spectral weight of
${\rm Im}P({\bf k}_{\rm F},{\bf p}-{\bf k}_{\rm F},{\bf k'}_{\rm F},\omega)$
exhibits a largest value at around the antinodal region, however, the most striking feature
is that the actual minimum of the spectral weight of
${\rm Im}P({\bf k}_{\rm F},{\bf p}-{\bf k}_{\rm F},{\bf k'}_{\rm F},\omega)$ does not appear
at around the node, but locates exactly at the tips of the Fermi arcs. This special angular
dependence of ${\rm Im}P({\bf k}_{\rm F},{\bf p}-{\bf k}_{\rm F},{\bf k'}_{\rm F},\omega)$
therefore induces an EFS reconstruction to form the Fermi arcs as shown in Fig. \ref{EFS-map}
with almost all the spectral weight of the electron excitation spectrum that resides at
around the tips of the Fermi arcs.

The result shown in Fig. \ref{effective-spin-propagator} therefore indicates that the main
contribution in $P({\bf k},{\bf p}-{\bf k},{\bf k'},\omega)$ comes from the part of the
momentum ${\bf p}={\bf k}$. In this case, the term
$\Phi({\bf k'})-\Phi({\bf k'}+{\bf k}-{\bf p})\sim 0$ in the right-hand side of
Eq. (\ref{electron-collision-2}), and then
$\delta(\bar{\varepsilon}_{\bf k}+\bar{\varepsilon}_{\bf k'}-\bar{\varepsilon}_{{\bf p}
+{\bf G}}-\bar{\varepsilon}_{{\bf k}'+{\bf k}-{\bf p}})$ in the right-hand side of
Eq. (\ref{electron-collision-2}) can be replaced by the integral identity as \cite{Lee21},
\begin{widetext}
\begin{eqnarray}\label{identity}
\delta(\bar{\varepsilon}_{\bf k}&+&\bar{\varepsilon}_{\bf k'}
-\bar{\varepsilon}_{{\bf p}+{\bf G}}-\bar{\varepsilon}_{{\bf k}'+{\bf k}-{\bf p}})
=\int^{\infty}_{-\infty}d\omega\delta(\bar{\varepsilon}_{\bf k}
-\bar{\varepsilon}_{{\bf p}+{\bf G}}-\omega)\delta(\omega+\bar{\varepsilon}_{\bf k'}
-\bar{\varepsilon}_{{\bf k}'+{\bf k}-{\bf p}}).~~~~~
\end{eqnarray}
\end{widetext}
On the other hand, the umklapp scattering process occurs mainly at around EFS, i.e.,
${\bf k'}\approx {\bf k'}_{\rm F}$, therefore the momentum ${\bf k'}$ in the effective
spin propagator $P({\bf k},{\bf p}-{\bf k},{\bf k}',\omega)$ can be approximately replaced
by the reduced effective spin propagator $\bar{P}({\bf k},{\bf p}-{\bf k},\omega)$,
\begin{eqnarray}\label{ESP-5}
\bar{P}({\bf k},{\bf p}-{\bf k},\omega)={1\over W_{\rm sp}}
P({\bf k},{\bf p}-{\bf k},{\bf k'}_{\rm F},\omega), ~~~~~~
\end{eqnarray}
where following the common practice, the scattering probability for two electrons has been
normalized with the normalization factor $W^{2}_{\rm sp}=(1/N^{2})\sum_{{\bf k},{\bf p}}
\int |{\rm Im}\bar{P}({\bf k},{\bf p}-{\bf k},\omega)|^{2}d\omega$. Substituting above
results in Eqs. (\ref{identity}) and (\ref{ESP-5}) into Eq. (\ref{electron-collision-2}),
the electron-electron collision in Eq. (\ref{electron-collision-2}) can be expressed
explicitly as \cite{Lee21},
\begin{widetext}
\begin{eqnarray}\label{electron-collision-3}
I_{\rm e-e}&=&{1\over N^{2}}\sum_{{\bf k}',{\bf p}} {2\over T}|\bar{P}({\bf k},{\bf p}
-{\bf k},\bar{\varepsilon}_{\bf k}-\bar{\varepsilon}_{{\bf p}+{\bf G}})|^{2}
[\tilde{\Phi}({\bf k})-\tilde{\Phi}({\bf p}+{\bf G})]n_{\rm F}(\bar{\varepsilon}_{\bf k})
n_{\rm F}(\bar{\varepsilon}_{{\bf k}'})[1-n_{\rm F}(\bar{\varepsilon}_{{\bf p}+{\bf G}})]
[1-n_{\rm F}(\bar{\varepsilon}_{{\bf k}'+{\bf k}-{\bf p}})]\nonumber\\
&\times& \int^{\infty}_{-\infty}d\omega\delta(\bar{\varepsilon}_{\bf k}
-\bar{\varepsilon}_{{\bf p}+{\bf G}}-\omega)
\delta(\omega+\bar{\varepsilon}_{\bf k'}-\bar{\varepsilon}_{{\bf k}'+{\bf k}-{\bf p}})
\nonumber\\
&=&{1\over N^{2}}\sum_{{\bf k}',{\bf p}} {2\over T}|\bar{P}({\bf k},{\bf p},
\bar{\varepsilon}_{\bf k}-\bar{\varepsilon}_{{\bf p}+{\bf k}+{\bf G}})|^{2}
[\tilde{\Phi}({\bf k})-\tilde{\Phi}({\bf p}+{\bf k}+{\bf G})]
n_{\rm F}(\bar{\varepsilon}_{\bf k})
n_{\rm F}(\bar{\varepsilon}_{{\bf k}'})
[1-n_{\rm F}(\bar{\varepsilon}_{{\bf p}+{\bf k}+{\bf G}})]
[1-n_{\rm F}(\bar{\varepsilon}_{{\bf k}'-{\bf p}})]\nonumber\\
&\times& \int^{\infty}_{-\infty}d\omega\delta(\bar{\varepsilon}_{\bf k}
-\bar{\varepsilon}_{{\bf p}+{\bf k}+{\bf G}}-\omega)
\delta(\omega+\bar{\varepsilon}_{\bf k'}-\bar{\varepsilon}_{{\bf k}'-{\bf p}}).
\end{eqnarray}
\end{widetext}
Now we replace the momentum ${\bf k}'$ integration by an integration along EFS and one
perpendicular to it, i.e.,
$(1/N)\sum_{{\bf k}'}=\int {\rm k}'d{\rm k}'d\theta_{{\rm k}'}/(2\pi)^{2}$, where the
$\theta_{{\rm k}'}$ specifies a patch of EFS in the direction $\theta_{{\rm k}'}$ as shown
in Fig. \ref{scatter-process}, and then the radial integration $\int d{\rm k}'$ is replaced
by an integral over $\int d{\rm k}'=\int d\bar{\varepsilon}_{{\bf k}'}/{\rm v}_{\rm F}$.
In this case, the above electron-electron collision in Eq. (\ref{electron-collision-3}) can
be simplified as,
\begin{widetext}
\begin{eqnarray}\label{electron-collision-4}
I_{\rm e-e}&=&{1\over 2\pi}{2{\rm k}_{\rm F}\over T{\rm v}^{2}_{\rm F}}{1\over N}
\sum_{{\bf p}}{1\over |{\bf p}|}\int {d\omega\over 2\pi}
|\bar{P}({\bf k},{\bf p},\bar{\varepsilon}_{\bf k}
-\bar{\varepsilon}_{{\bf p}+{\bf k}+{\bf G}})|^{2}[\tilde{\Phi}({\bf k})
-\tilde{\Phi}({\bf p}+{\bf k}+{\bf G})]n_{\rm F}(\bar{\varepsilon}_{\bf k})
[1-n_{\rm F}(\bar{\varepsilon}_{{\bf p}+{\bf k}+{\bf G}})]\nonumber\\
&\times& \omega[1+n_{\rm B}(\omega)]
\delta(\bar{\varepsilon}_{\bf k}-\bar{\varepsilon}_{{\bf p}+{\bf k}+{\bf G}}-\omega).
\end{eqnarray}
\end{widetext}
For the obtain of the above equation (\ref{electron-collision-4}), the following identity,
\begin{eqnarray}\label{identity-1}
\int_{-\infty}^{+\infty}d\varepsilon n_{\rm F}(\varepsilon-\omega)[1-n_{\rm F}(\varepsilon)]
=\omega[1+n_{\rm B}(\omega)],~~~~
\end{eqnarray}
has been used, where the appearance of the boson distribution function $n_{\rm B}(\omega)$
in the right-hand side signals that we are describing a particle-hole effective
spin excitation which has the boson statistics \cite{Lee21}.

Now we turn to evaluate the momentum ${\bf p}$ integration, which is quite similar to the
evaluation of the momentum ${\bf k}'$ integration in Eqs. (\ref{electron-collision-3}) and
(\ref{electron-collision-4}). After a straightforward calculation for the momentum ${\bf p}$
integration in Eq. (\ref{electron-collision-4}), the electron-electron collision term can be
obtained explicitly as
\begin{widetext}
\begin{eqnarray}\label{electron-collision-5}
I_{\rm e-e}&=&{1\over (2\pi)^{2}}{2{\rm k}^{2}_{\rm F}\over T{\rm v}^{3}_{\rm F}}
\int {d\theta'\over 2\pi}\int {d\omega\over 2\pi}{1\over {\rm p}(\theta,\theta')}
|\bar{P}[k(\theta),p(\theta,\theta'),\omega]|^{2}[\Phi(\theta)-\Phi(\theta')]
n_{\rm F}(\bar{\varepsilon}_{k(\theta)})
[1-n_{\rm F}(\bar{\varepsilon}_{k(\theta)}-\omega)]\omega[1+n_{\rm B}(\omega)],~~~~~
\end{eqnarray}
where $\tilde{\Phi}({\bf k})$ and $\tilde{\Phi}({\bf p}+{\bf k}+{\bf G})$ in the right-hand
of side have been replaced by $\Phi(\theta)$ and $\Phi(\theta')$, respectively, and at
low-energy regime, the Boltzmann equation in Eq. (\ref{Boltzmann-equation-3}) can
be expressed as,
\begin{eqnarray}\label{Boltzmann-equation-4}
e{\bf v}_{\rm F}\cdot{\bf E}{\partial n_{\rm F}(\bar{\varepsilon}_{k(\theta)})\over
\partial\bar{\varepsilon}_{k(\theta)}}&=&{1\over (2\pi)^{2}}{2{\rm k}^{2}_{\rm F}\over
T{\rm v}^{3}_{\rm F}}\int {d\theta'\over 2\pi}\int {d\omega\over 2\pi}
{1\over {\rm p}(\theta,\theta')}|\bar{P}[k(\theta),p(\theta,\theta'),\omega]|^{2}
[\Phi(\theta)-\Phi(\theta')]\nonumber\\
&\times& n_{\rm F}(\bar{\varepsilon}_{k(\theta)})
[1-n_{\rm F}(\bar{\varepsilon}_{k(\theta)}-\omega)]\omega[1+n_{\rm B}(\omega)].~~~~~
\end{eqnarray}
Integrating both the left-hand and right-hand sides over the energy
$\bar{\varepsilon}_{k(\theta)}$, the Boltzmann equation in Eq. (\ref{Boltzmann-equation-4})
can be obtained explicitly as,
\begin{eqnarray}\label{Boltzmann-equation-5}
e{\bf v}_{\rm F}\cdot{\bf E}&=&-{1\over (2\pi)^{2}}{2{\rm k}^{2}_{\rm F}\over
T{\rm v}^{3}_{\rm F}}\int {d\theta'\over 2\pi}\int {d\omega\over 2\pi}
{1\over {\rm p}(\theta,\theta')}|\bar{P}[k(\theta),p(\theta,\theta'),\omega]|^{2}
[\Phi(\theta)-\Phi(\theta')]\omega^{2}n_{\rm B}(\omega)[1+n_{\rm B}(\omega)]\nonumber\\
&=&-2\int {d\theta'\over {2\pi}}\zeta(\theta')F(\theta,\theta')[\Phi(\theta)-\Phi(\theta')],
\end{eqnarray}
which is the same as quoted in Eq. (\ref{electron-collision}) of the main text.

\end{widetext}

\section{Derivation of function $I_{jj'}(\theta,\theta',{\bf q},{\bf q}')$}
\label{reduced-kernel-function}

In this Appendix, we sketch the derivation of the function
$I_{jj'}(\theta,\theta',{\bf q},{\bf q}')$ in Eq. (\ref{C-function-1}) of the main text.
For a convenience in the following discussions, we define the functions:
$C=C_{j}(\theta,\theta',{\bf q})$, $C'=C_{j'}(\theta,\theta',{\bf q}')$, and
$I(C,C')=I_{jj'}(\theta,\theta',{\bf q},{\bf q}')$. In this case, the function
$I_{jj'}(\theta,\theta',{\bf q},{\bf q}')$ in Eq. (\ref{C-function-1}) of the main text
can be rewritten in the complex-plane as,
\begin{equation}\label{C-function-5}
I_{z}(C,C')=\int{dz\over i2\pi}{e^{z}\over (e^{z}-1)^{2}}{z^{2}
\over [z^{2}-C^{2}][z^{2}-C'^{2}]},
\end{equation}
with the closed integration path in the upper complex-plane, and then this integration along
the real axis can be obtained as,
\begin{widetext}
\begin{eqnarray}\label{C-function-6}
I(C,C')&=&P\int_{-\infty}^{\infty}{dx\over i2\pi}{e^{x}\over (e^{x}-1)^{2}}{x^{2}\over
[x^{2}-C^{2}][x^{2}-C'^{2}]}= res[z=i\omega_{n}=i2\pi n(n\in N^{+})]\nonumber\\
&+&{1\over 2}\sum\limits_{z_{0}}\lim_{z\to z_{0}}(z-z_{0})
{e^{z}\over (e^{z} -1)^2}{z^{2}\over [z^{2}-C^{2}][z^2 - C'^{2}]},~~~~~
\end{eqnarray}
where $z_{0}=i\omega_{n}$, $z_{0}=\pm C$, and $z_{0}=\pm C'$ are the poles. With the help
of the above results in Eq. (\ref{C-function-6}), the function $I(C,C')$ in
Eq. (\ref{C-function-1}) can be expressed in terms of the residues of the corresponding
poles as,
\begin{eqnarray}\label{C-function-7}
I(C,C')&=&P\int_{-\infty}^{\infty}{dx\over i2\pi}
{e^{x}\over (e^{x}-1)^{2}}{x^{2}\over [x^{2}-C^{2}][x^{2}-C'^{2}]}
= -2\sum\limits_{n>0}{i\omega_{n}[i\omega_{n}^{4}-C^{2}C'^{2}]\over [i\omega_{n}^{2}
-C^{2}]^{2}[i\omega_{n}^{2}-C'^{2}]^{2}}\nonumber\\
&=&-\sum\limits_{n}{i|\omega_{n}|[i\omega_{n}^{4}-C^{2}C'^{2}]\over [i\omega_{n}^{2}
-C^{2}]^{2}[i\omega_{n}^{2}-C'^{2}]^{2}}.~~~~~
\end{eqnarray}
For obtaining the above result in Eq. (\ref{C-function-7}), the following identities,
\begin{subequations}\label{residues-1}
\begin{eqnarray}
\lim_{z\to 0}z{e^{z}\over (e^{z} -1 )^{2}}{z^{2}\over [z^{2}-C^{2}][z^{2}- C'^{2}]} &=& 0,
~~~~~ \\
\lim_{z\to \pm C}(z-\pm C){e^{z}\over (e^{z}-1)^{2}}{z^{2}
\over [z^{2}-C^{2}][z^{2}-C'^{2}]}
&=& \pm {1\over 2}{e^{C}\over [e^{C}-1]^{2}}{C\over C^{2}-C'^{2}}, \\
\lim_{z\to \pm C'}(z-\pm C'){e^{z}\over (e^{z}-1)^{2}}{z^{2}
\over [z^{2}-C^{2}][z^{2}-C'^{2}]}
&=& \pm {1\over 2}{e^{C'}\over [e^{C'}-1]^{2}}{C'\over C'^{2}-C^{2}},\\
res[z=i\omega_{n}=i2\pi n(n\in N^{+})]&=& -2{i\omega_{n}[i\omega_{n}^{4}-C^{2}C'^{2}]
\over [i\omega_{n}^{2}-C^{2}]^{2}[i\omega_{n}^{2}-C'^{2}]^{2}}, ~~~~
\end{eqnarray}
\end{subequations}
have been used. We now introduce the following integration,
\begin{eqnarray}
\bar{I}_{z}(C,C')=\int {dz\over i2\pi}{|z|[z^{4}-C^{2}C'^{2}]\over [z^{2}-C^{2}]^{2}
[z^{2}-C'^{2}]^{2}}n_{B}(z),
\end{eqnarray}
with the integration path along the closed path at infinity in the complex-plane, which
can be derived directly as,
\begin{eqnarray}\label{Cz-function-7}
\bar{I}_{z}(C,C')=\sum\limits_{n}{|\omega_n|[i\omega_{n}^{4}-C^{2}C'^{2}]
\over [i\omega_{n}^{2}-
C^{2}]^{2}[i\omega_{n}^{2}-C'^{2}]^{2}}
+\sum\limits_{z_{0}}\lim_{z\to z_{0}}{d\over dz}(z-z_{0})^{2}{|z|[z^{4}-C^{2}C'^{2}]
\over [z^{2}-C^{2}]^{2}[z^{2}-C'^{2}]^{2}}n_{B}(z)=0,~~~~~
\end{eqnarray}
where the poles are located at $z_{0}=\pm C$ and $z_{0}=\pm C'$, while the derivative in
the above equation (\ref{Cz-function-7}) can be calculated straightforwardly as,
\begin{eqnarray}\label{residues-5}
&&{d\over dz}(z-z_{0})^{2}{|z|[z^{4}-C^{2}C'^{2}]\over [z^{2}-C^{2}]^{2}[z^{2}-C'^{2}]^{2}}
n_{B}(z)|_{z\to z_{0}} = {d\over dz}{|z|[z^{4}-C^{2}C'^{2}]\over
[z+z_{0}]^{2}[z^{2}-z_{0}'^{2}]^{2}}n_{B}(z)|_{z\to z_{0}}\nonumber\\
&&={z_{0}^{2}(z_{0}^{2}-z_{0}'^{2})^{2}[|z_{0}|'(z_{0}^{4}-C^{2}C'^{2})+4|z_{0}|z_{0}^{3}]
-|z_{0}|z_{0}(z_{0}^{2}-z_{0}'^{2})(z_{0}^{4}-C^{2}C'^{2})(5z_{0}^{2}-z_{0}'^{2})\over
4z_{0}^{4}(z_{0}^{2}-z_{0}'^{2})^{4}}{1\over e^{{z}_{0}}-1}\nonumber\\
&&- {|z_{0}|(z_{0}^{4}-C^{2}C'^{2})\over 4z_{0}^{2}(z_{0}^{2}-z_{0}'^{2})^{2}}
{e^{z_{0}}\over (e^{z_{0}}-1)^{2}},~~~~
\end{eqnarray}
and then the residues in Eq. (\ref{Cz-function-7}) corresponding to the poles at
$z_{0}=\pm C$ and $z_{0}=\pm C'$ are derived explicitly as,
\begin{subequations}\label{residues-6}
\begin{eqnarray}
res(z_{0}=C) &=& res(z_{0}=-C) ={C\over 4(C^{2}-C'^{2})}{e^{C}\over (e^{C}-1)^{2}},\\
res(z_{0}=C')&=& res(z_{0}=-C')={C'\over 4(C'^{2}-C^{2})}{e^{C'}\over (e^{C'}-1)^{2}},~~~~~~
\end{eqnarray}
\end{subequations}
respectively. With the help of the above results of the residues, the following identity is
obtained explicitly,
\begin{eqnarray}\label{residues-7}
-\sum\limits_{n}{|\omega_{n}|(i\omega_{n}^{4}-C^{2}C'^{2})\over (i\omega_{n}^{2}
-C^{2})^{2}(i\omega_{n}^{2}-C'^{2})^{2}}
= {1\over 2(C^{2}-C'^{2})}\left [ {Ce^{C}\over (e^{C}-1)^{2}}-{C'e^{C'}
\over (e^{C'}-1)^{2}} \right]. ~~~~~~
\end{eqnarray}
Substituting the above result in Eq. (\ref{residues-7}) into Eq. (\ref{C-function-6}), the
function $I(C,C')$ can be obtained explicitly as,
\begin{equation}
I(C,C')={i\over 2[C^{2}-C'^{2}]}\left ( {Ce^{C}\over [e^{C}-1]^{2}}-{C'e^{C'}\over
[e^{C'}-1]^{2}}\right ),
\end{equation}
which is the same as quoted in Eq. (\ref{C-function-2}) of the main text.

\end{widetext}

\end{appendix}



\begin{thebibliography}{00}

\bibitem{Fujita12} See, e.g., the review, M. Fujita, H. Hiraka, M. Matsuda, M. Matsuura, J. M.
Tranquada, S. Wakimoto, G. Xu, and K. Yamada, Progress in neutron scattering studies of
spin excitations in high-$T_c$ cuprates, J. Phys. Soc. Jpan. {\bf 81}, 011007 (2012).

\bibitem{Bednorz86} J. G. Bednorz and K. A. M\"uller, Possible high $T_c$ superconductivity
in the Ba-La-Cu-O system, Z. Phys. B {\bf 64}, 189 (1986).

\bibitem{Vishik18} See, e.g., the review, I. M. Vishik, Photoemission perspective on 
pseudogap, superconducting fluctuations, and charge order in cuprates: a review of recent 
progress, Rep. Prog. Phys. {\bf 81}, 062501 (2018).

\bibitem{Campuzano04} See, e.g., the review, J. C. Campuzano, M. R. Norman, M. Randeira,
Photoemission in the high-$T_c$ superconductors, in {\it Physics of Superconductors}, 
vol. II, edited by K. H. Bennemann and J. B. Ketterson (Springer, Berlin Heidelberg New 
York, 2004), p. 167.

\bibitem{Damascelli03} See, e.g., the review, A. Damascelli, Z. Hussain, and Z.-X. Shen,
Angle-resolved photoemission studies of the cuprate superconductors, Rev. Mod. Phys. 
{\bf 75}, 473 (2003).

\bibitem{Fink07} See, e.g., the review, J. Fink, S. Borisenko, A. Kordyuk, A. Koitzsch, J.
Geck, V. Zabalotnyy, M. Knupfer, B. B\"uechner, and H. Berger, Dressing of the charge 
carriers in high-$T_c$ superconductors, in {\it Lecture Notes in Physics}, vol. 715, 
edited by S. H\"ufner (Springer-Verlag Berlin Heidelberg, 2007), p. 295.

\bibitem{Keimer15} See, e.g., the review, B. Keimer, S. A. Kivelson, M. R. Norman, S. 
Uchida, and J. Zaanen, From quantum matter to high-temperature superconductivity in 
copper oxides, Nature {\bf 518}, 179 (2015).

\bibitem{Hussey08} See, e.g., the review, N. E. Hussey, Phenomenology of the normal state 
in-plane transport properties of high-$T_c$ cuprates, J. Phys.: Condens. Matter {\bf 20}, 
123201 (2008).

\bibitem{Timusk99} See, e.g., the review, T. Timusk and B. Statt, The pseudogap in 
high-temperature superconductors: an experimental survey, Rep. Prog. Phys. {\bf 62}, 61 
(1999).

\bibitem{Kastner98} See, e.g., the review, M. A. Kastner, R. J. Birgeneau, G. Shirane, 
and Y. Endoh, Magnetic, transport, and optical properties of monolayer copper oxides, 
Rev. Mod. Phys. {\bf 70}, 897 (1998).

\bibitem{Ando04} Y. Ando, S. Komiya, K. Segawa, S. Ono, and Y. Kurita, Electronic phase 
diagram of high-$T_c$ cuprate superconductors from a mapping of the in-plane resistivity 
curvature, Phys. Rev. Lett. {\bf 93}, 267001 (2004).

\bibitem{Barisic13} N. Bari\v{s}i\'c, M. K. Chan, Y. Li, G. Yu, X. Zhao, M. Dressel, A.
Smontara, and M. Greven, Universal sheet resistance and revised phase diagram of the 
cuprate high-temperature superconductors, Proc. Natl. Acad. Sci. USA {\bf 110}, 12235 
(2013).

\bibitem{Pelc20} D. Pelc, M. J. Veit, C. J. Dorow, Y. Ge, N. Bari\v{s}i\'c, and M. Greven, 
Resistivity phase diagram of cuprates revisited, Phys. Rev. B {\bf 102}, 075114 (2020).

\bibitem{Schrieffer64} See, e.g., J. R. Schrieffer, {\it Theory of Superconductivity}, 
Benjamin, New York, 1964.

\bibitem{Abrikosov88} See, e.g., A. A. Abrikosov, {\it Fundamentals of the Theory of 
Metals}, Elsevier Science Publishers B. V., 1988.

\bibitem{Mahan81} See, e.g., G. D. Mahan, {\it Many-Particle Physics}, (Plenum Press, New 
York, 1981).

\bibitem {Allen89} See, e.g., the review, P. B. Allen, Z. Fisk, and A. Migliori, Normal 
state transport and elastic properties of high $T_c$ materials and related compounds, in
{\it Physical Properties of High Temperature Superconductors} I, edited by D. M. Ginsberg
(World Scientific, Singapore, 1989), p. 213.

\bibitem{Gurvitch87} M. Gurvitch and A. T. Fiory, Resistivity of 
La$_{1.825}$Sr$_{0.175}$CuO$_{4}$ and YBa$_{2}$Cu$_{3}$O$_{7}$ to 1100 K: absence of 
saturation and its implications, Phys. Rev. Lett. {\bf 59}, 1337 (1987).

\bibitem{Takagi92} H. Takagi, B. Batlogg, H. L. Kao, J. Kwo, R. J. Cava, J. J. Krajewski, 
and W. F. Peck, Jr., Systematic evolution of temperature-dependent resistivity in 
La$_{2-x}$Sr$_{x}$CuO$_{4}$, Phys. Rev. Lett. {\bf 69}, 2975 (1992).

\bibitem{Martin90} S. Martin, A. T. Fiory, R. M. Fleming, L. F. Schneemeyer, and J. V. 
Waszczak, Normal-state transport properties of Bi$_{2+x}$Sr$_{2-y}$Cu$_{6\pm \delta}$ 
crystals, Phys. Rev. B {\bf 41}, 846 (1990).

\bibitem{Mandrus92} D. Mandrus, L. Forro, C. Kendziora, and L. Mihaly, Resistivity study 
of Bi$_{2}$Sr$_{2}$Ca$_{1-x}$Y$_{x}$Cu$_{2}$O$_{8}$ single crystals, Phys. Rev. B {\bf 45},
12640 (1992).

\bibitem{Ando01} Y. Ando, A. N. Lavrov, S. Komiya, K. Segawa, and X. F. Sun, Mobility of 
the doped holes and the antiferromagnetic correlations in underdoped high-$T_c$ cuprates, 
Phys. Rev. Lett. {\bf 87}, 017001 (2001).

\bibitem{Daou09} R. Daou, N. Doiron-Leyraud, D. LeBoeuf, S. Y. Li, F. Lalibert\'e, O.
Cyr-Choini\'ere, Y. J. Jo, L. Balicas, J.-Q. Yan, J.-S. Zhou, J. B. Goodenough, and L.
Taillefer, Linear temperature dependence of resistivity and change in the Fermi surface
at the pseudogap critical point of a high-$T_c$ superconductor, Nat. Phys. {\bf 5}, 31 
(2009).

\bibitem{Cooper09} R. A. Cooper, Y. Wang, B. Vignolle, O. J. Lipscombe, S. M. Hayden, Y.
Tanabe, T. Adachi, Y. Koike, M. Nohara, H. Takagi, C. Proust, N. E. Hussey, Anomalous
criticality in the electrical resistivity of La$_{2-x}$Sr$_{x}$CuO$_{4}$, Science 
{\bf 323}, 603 (2009).

\bibitem{Legros19} A. Legros, S. Benhabib, W. Tabis, F. Lalibert\'e, M. Dion, M. Lizaire, 
B. Vignolle, D. Vignolles, H. Raffy, Z. Z. Li, P. Auban-Senzier, N. Doiron-Leyraud, P. 
Fournier, D. Colson, L. Taillefer, and C. Proust, Universal $T$-linear resistivity and 
Planckian dissipation in overdoped cuprates, Nat. Phys. {\bf 15}, 142 (2019).

\bibitem{Yuan22} J. Yuan, Q. Chen, K. Jiang, Z. Feng, Z. Lin, H. Yu, G. He, J. Zhang, X.
Jiang, X. Zhang, Y. Shi, Y. Zhang, M. Qin, Z. Cheng, N. Tamura, Y.-F. Yang, T. Xiang, J.
Hu, I. Takeuchi, K. Jin, and Z. Zhao, Scaling of the strange-metal scattering in
unconventional superconductors, Nature {\bf 602}, 431 (2022).

\bibitem{Ayres21} J. Ayres, M. Berben, M. \v{C}ulo, Y.-T. Hsu, E. van Heumen, Y. Huang, J.
Zaanen, T. Kondo, T. Takeuchi, J. R. Cooper, C. Putzke, S. Friedemann, A. Carrington, and
N. E. Hussey, Incoherent transport across the strange-metal regime of overdoped
cuprates, Nature {\bf 595}, 661 (2021).

\bibitem{Grisso21} G. Grissonnanche, Y. Fang, A. Legros, S. Verret, F. Lalibert\'e, C.
Collignon, J. Zhou, D. Graf, P. A. Goddard, L. Taillefer, and B. J. Ramshaw, Linear-in
temperature resistivity from an isotropic Planckian scattering rate, Nature
{\bf 595}, 667 (2021).

\bibitem{Varma89} C. M. Varma, P. B. Littlewood, S. Schmitt-Rink, E. Abrahams, and A. E.
Ruckenstein, Phenomenology of the normal state of Cu-O high-temperature superconductors,
Phys. Rev. Lett. {\bf 63}, 1996 (1989).

\bibitem{Varma16} See, e.g., the review, C. M Varma, Quantum-critical fluctuations in 2D 
metals: strange metals and superconductivity in antiferromagnets and in cuprates, Rep. 
Prog. Phys. {\bf 79} 082501 (2016)

\bibitem{Varma20} See, e.g., the review, C. M. Varma, Colloquium: Linear in temperature 
resistivity and associated mysteries including high temperature superconductivity, Rev. 
Mod. Phys. {\bf 92}, 031001 (2020).

\bibitem{Damle97} K. Damle and S. Sachdev, Nonzero-temperature transport near quantum 
critical points, Phys. Rev. B {\bf 56}, 8714 (1997).

\bibitem{Sachdev11} S. Sachdev, {\it Quantum Phase Transitions}, (Cambridge University 
Press, 1999).

\bibitem{Zaanen04} J. Zaanen, Why the temperature is high, Nature {\bf 430}, 512 (2004).

\bibitem{Luca07} L. Dell'Anna and W. Metzner, Electrical resistivity near Pomeranchuk
instability in two dimensions, Phys. Rev. Lett. {\bf 98}, 136402 (2007).

\bibitem{Haldane18} F. D. M. Haldane, Fermi-surface geometry and “Planckian dissipation”,
arXiv:1811.12120.

\bibitem{Zaanen19} See, e.g., the review, J. Zaanen, Planckian dissipation, minimal
viscosity and the transport in cuprate strange metals, SciPost Phys. {\bf 6}, 061 (2019).

\bibitem{Hartnoll22} See, e.g., the review, S. A. Hartnoll and A. P. Mackenzie, Colloquium:
Planckian dissipation in metals, Rev. Mod. Phys. {\bf 94}, 041002 (2022).

\bibitem{Hussey03} N. E. Hussey, The normal state scattering rate in high-$T_c$ cuprates,
Eur. Phys. J. B {\bf 31}, 495 (2003).

\bibitem{Rice17} T. M. Rice, N. J. Robinson, and A. M. Tsvelik, Umklapp scattering as the
origin of $T$-linear resistivity in the normal state of high-$T_c$ cuprate superconductors,
Phys. Rev. B {\bf 96}, 220502(R) (2017).

\bibitem{Lee21} P. A. Lee, Low-temperature $T$-linear resistivity due to umklapp scattering
from a critical mode, Phys. Rev. B {\bf 104}, 035140 (2021).

\bibitem{Honerkamp01} C. Honerkamp, M. Salmhofer, N. Furukawa, and T. M. Rice, Breakdown
of the Landau-Fermi liquid in two dimensions due to umklapp scattering, Phys. Rev. B
{\bf 63}, 035109 (2001).

\bibitem{Hartnoll12} S. A. Hartnoll and D. M. Hofman, Locally critical resistivities from
umklapp scattering, Phys. Rev. Lett. {\bf 108}, 241601
(2012).

\bibitem{Tabis21} W. Tabi\'{s}, P. Pop\v{c}evi\'{c}, B. Klebel-Knobloch, I. Bialo, C. M. N.
Kumar, B. Vignolle, M. Greven, N. Bari\v{s}i\'{c}, Arc-to-pocket transition and quantitative
understanding of transport properties in cuprate superconductors, arXiv:2106.07457.

\bibitem{Liu21} Y. Liu, Y. Lan, and S. Feng, Peak structure in the self-energy of cuprate
superconductors, Phys. Rev. B {\bf 103}, 024525 (2021); S. Tan, Y. Liu, Y. Mou, and S. Feng,
Anisotropic dressing of electrons in electron-doped cuprate superconductors, Phys. Rev. B
{\bf 103}, 014503 (2021).

\bibitem{Cao21} Z. Cao, X. Ma, Y. Liu, H. Guo, and S. Feng, Characteristic energy of the
nematic-order state and its connection to enhancement of superconductivity in cuprate
superconductors, Phys. Rev. B {\bf 104}, 224503 (2021); Z. Cao, Y. Liu, H. Guo, and S. 
Feng, Enhancement of superconductivity by electronic nematicity in cuprate superconductors, 
Phil. Mag. {\bf 102}, 918 (2022).

\bibitem{Zeng22} M. Zeng, X. Li, Y. Wang, and S. Feng, Influence of impurities on the 
electronic structure in cuprate superconductors, Phys. Rev. B {\bf 106}, 054512 (2022).

\bibitem {Feng16} S. Feng, D. Gao, and H. Zhao, Charge order driven by Fermi-arc instability 
and its connection with pseudogap in cuprate superconductors, Phil. Mag.  {\bf 96}, 1245 
(2016); H. Zhao, D. Gao, and S. Feng, Pseudogap-generated a coexistence of Fermi arcs and 
Fermi pockets in cuprate superconductors, Physica C {\bf 534}, 1 (2017); X. Ma, Z. Cao, and 
S. Feng, unpublished.

\bibitem{Anderson87} P. W. Anderson, The resonating valence bond state in La$_{2}$CuO$_{4}$
and superconductivity, Science {\bf 235}, 1196 (1987).

\bibitem {Yu92} See, e.g., the review, L. Yu, Many-body problems in high temperature 
superconductors, in {\it Recent Progress in Many-Body Theories}, edited by T. L. Ainsworth, 
C. E. Campbell, B. E. Clements, and E. Krotscheck (Plenum, New York, 1992), Vol. {\bf 3}, 
p. 157.

\bibitem {Lee06} See, e.g., the review, P. A. Lee, N. Nagaosa, and X.-G. Wen, Doping a Mott 
insulator: Physics of high-temperature superconductivity, Rev. Mod. Phys. {\bf 78}, 17 (2006).

\bibitem{Edegger07} See, e.g., the review, B. Edegger, V. N. Muthukumar, and C. Gros, 
Gutzwiller–RVB theory of high-temperature superconductivity: Results from renormalized 
mean-field theory and variational Monte Carlo calculations, Adv. Phys. {\bf 56}, 927 (2007).

\bibitem {Spalek22} J. Spalek, M. Fidrysiak, M. Zegrodnik, and A. Biborski, Superconductivity 
in high-$T_c$ and related strongly correlated systems from variational perspective: Beyond 
mean field theory, Phys. Rep. {\bf 959}, 1 (2022).

\bibitem {Zhang93} L. Zhang, J. K. Jain, and V. J. Emery, Importance of the local constraint 
in slave-boson theories, Phys. Rev. B {\bf 47}, 3368 (1993).

\bibitem {Feng0494} S. Feng, J. Qin, and T. Ma, A gauge invariant dressed holon and spinon
description of the normal state of underdoped cuprates, J. Phys.: Condens. Matter {\bf 16}, 
343 (2004); S. Feng, Z. B. Su, and L. Yu, Fermion-spin transformation to implement the 
charge-spin separation, Phys. Rev. B {\bf 49}, 2368 (1994).

\bibitem {Feng15} See, e.g., the review, S. Feng, Y. Lan, H. Zhao, L. Kuang, L. Qin, and
X. Ma, Kinetic-energy-driven superconductivity in cuprate superconductors, Int. J. Mod.
Phys. B {\bf 29}, 1530009 (2015).

\bibitem{Feng0306} S. Feng, Kinetic energy driven superconductivity in doped cuprates,
Phys. Rev. B {\bf 68}, 184501 (2003); S. Feng, T. Ma, and H. Guo, Magnetic nature of
superconductivity in doped cuprates, Physica C {\bf 436}, 14 (2006).

\bibitem{Feng12} S. Feng, H. Zhao, and Z. Huang, Two gaps with one energy scale in cuprate
superconductors, Phys. Rev. B. {\bf 85}, 054509 (2012); Phys. Rev. B {\bf 85}, 099902(E) 
(2012).

\bibitem{Feng15a} S. Feng, L. Kuang, and H. Zhao, Electronic structure of cuprate 
superconductors in a full charge-spin recombination scheme, Physica C {\bf 517}, 5 (2015).

\bibitem{Scalapino86} D. J. Scalapino, E. Loh, Jr., and J. E. Hirsch, $d$-wave pairing near 
a spin-density-wave instability, Phys. Rev. B {\bf 34}, 8190 (1986).

\bibitem{Miyake86} K. Miyake, S. Schmitt-Rink, and C. M. Varma, Spin-fluctuation-mediated
even-parity pairing in heavy-fermion superconductors, Phys. Rev. B {\bf 34}, 6554 (1986).

\bibitem{Monthoux91} P. Monthoux, A. V. Balatsky, and D. Pines, Toward a theory of
high-temperature superconductivity in the antiferromagnetically correlated cuprate oxides,
Phys. Rev. Lett. {\bf 67}, 3448 (1991); P. Monthoux and D. Pines, YBa$_{2}$Cu$_{3}$O$_{7}$:
A nearly antiferromagnetic Fermi liquid, Phys. Rev. B {\bf 47}, 6069 (1993).

\bibitem{Monthoux07} P. Monthoux, D. Pines, and G. G. Lonzarich, Superconductivity without
phonons, Nature {\bf 450}, 1177 (2007).

\bibitem{Eliashberg60} G. M. Eliashberg, Interactions between electrons and lattice 
vibrations in a superconductor, Sov. Phys. JETP {\bf 11}, 696 (1960).

\bibitem{Brinckmann01} J. Brinckmann and P. A. Lee, Renormalized mean-field theory of 
neutron scattering in cuprate superconductors, Phys. Rev. B {\bf 65}, 014502 (2001).

\bibitem{Restrepo23} F. Restrepo, J. Zhao, J. C. Campuzano, and U. Chatterjee, 
Temperature and carrier concentration dependence of Fermi arcs in moderately underdoped
Bi$_{2}$Sr$_{2}$CaCu$_{2}$O$_{8+\delta}$ cuprate high-temperature superconductors: A 
joint density of states perspective, Phys. Rev. B {\bf 107}, 174519 (2023).

\bibitem{Norman98} M. R. Norman, H. Ding, M. Randeria, J. C. Campuzano, T. Yokoya, T. 
Takeuchi, T. Takahashi, T. Mochiku, K. Kadowaki, P. Guptasarma, and D. G. Hinks, 
Destruction of the Fermi surface in underdoped high-$T_c$ superconductors, Nature 
{\bf 392}, 157 (1998).

\bibitem{Shi08} M. Shi, J. Chang, S. Pailh\'es, M. R. Norman, J. C. Campuzano, M. M\'ansson,
T. Claesson, O. Tjernberg, A. Bendounan, L. Patthey, N. Momono, M. Oda, M. Ido, C. Mudry,
and J. Mesot, Coherent $d$-wave superconducting gap in underdoped La$_{2-x}$Sr$_{x}$CuO$_{4}$
by angle-resolved photoemission spectroscopy, Phys. Rev. Lett. {\bf 101}, 047002 (2008).

\bibitem{Sassa11} Y. Sassa, M. Radovi\'c, M. M\'ansson, E. Razzoli, X. Y. Cui, S. Pailh\'es,
S. Guerrero, M. Shi, P. R. Willmott, F. Miletto Granozio, J. Mesot, M. R. Norman, and L.
Patthey, Ortho-II band folding in YBa$_{2}$Cu$_{3}$O$_{7-\delta}$ films revealed by 
angle-resolved photoemission, Phys. Rev. B {\bf 83}, 140511(R) (2011).

\bibitem{Meng11} J. Q. Meng, M. Brunner, K.-H. Kim, H.-G. Lee, S.-I. Lee, J. S. Wen, Z. J. 
Xu, G. D. Gu, and G.-H. Gweon, Momentum-space electronic structures and charge orders of 
the high-temperature superconductors Ca$_{2-x}$Na$_{x}$CuO$_{2}$Cl$_{2}$ and 
Bi$_{2}$Sr$_{2}$CaCu$_{2}$O$_{8+\delta}$, Phys. Rev. B {\bf 84}, 060513 (R) (2011).

\bibitem{Horio16} M. Horio, T. Adachi, Y. Mori, A. Takahashi, T. Yoshida, H. Suzuki, L. C.
C. Ambolode II, K. Okazaki, K. Ono, H. Kumigashira, H. Anzai, M. Arita, H. Namatame, M.
Taniguchi, D. Ootsuki, K. Sawada, M. Takahashi, T. Mizokawa, Y. Koike, and A. Fujimori,
Suppression of the antiferromagnetic pseudogap in the electron-doped high-temperature
superconductor by protect annealing, Nat. Commun. {\bf 7}, 10567 (2016).

\bibitem{Loret18} B. Loret, Y. Gallais, M. Cazayous, R. D. Zhong, J. Schneeloch, G. D. Gu,
A. Fedorov, T. K. Kim, S. V. Borisenko, and A. Sacuto, Raman and ARPES combined study on 
the connection between the existence of the pseudogap and the topology of the Fermi 
surface in Bi$_{2}$Sr$_{2}$CaCu$_{2}$O$_{8+\delta}$, Phys. Rev. B {\bf 97}, 174521 (2018).

\bibitem{Chen19} S. D. Chen, M. Hashimoto, Y. He, D. Song, K. J. Xu, J. F. He, T. P.
Devereaux, H. Eisaki, D. H. Lu, J. Zaanen, and Z. -X. Shen, Incoherent strange metal
sharply bounded by a critical doping in Bi2212, Science {\bf 366}, 1099 (2019).

\bibitem{Loret17} B. Loret, S. Sakai, S. Benhabib, Y. Gallais, M. Cazayous, M. A. M\'easson, 
R. D. Zhong, J. Schneeloch, G. D. Gu, A. Forget, D. Colson, I. Paul, M. Civelli, and A. 
Sacuto, Vertical temperature boundary of the pseudogap under the superconducting dome in 
the phase diagram of Bi$_{2}$Sr$_{2}$CaCu$_{2}$O$_{8+\delta}$, Phys. Rev. B {\bf 96}, 
094525 (2017).

\bibitem{Kaminski15} A. Kaminski, T. Kondo, T. Takeuchi, and G. Gu, Pairing, pseudogap and 
Fermi arcs in cuprates, Phil. Mag. {\bf 95}, 453 (2015).

\bibitem{Comin14} R. Comin, A. Frano, M. M. Yee, Y. Yoshida, H. Eisaki, E. Schierle, E.
Weschke, R. Sutarto, F. He, A. Soumyanarayanan, Yang He, M. L. Tacon, I. S. Elfimov,
Jennifer E. Hoffman, G. A. Sawatzky, B. Keimer, and A. Damascelli, Charge order driven
by Fermi-arc instability in Bi$_{2}$Sr$_{2-x}$La$_{x}$CuO$_{6+\delta}$, Science {\bf 343}, 
390 (2014).

\bibitem{Fujita14} K. Fujita, C. K. Kim, I. Lee, J. Lee, M. H. Hamidian, I. A. Firmo, S.
Mukhopadhyay, H. Eisaki, S. Uchida, M. J. Lawler, E.-A. Kim, and J. C. Davis, Simultaneous
 transitions in cuprate momentum-space topology and electronic symmetry breaking, Science
{\bf 344}, 612 (2014).

\bibitem{Hufner08} See, e.g., the review, S. H\"ufner, M. A. Hossain, A. Damascelli, and 
G. A. Sawatzky, Two gaps make a high-temperature superconductor?, Rep. Prog. Phys. {\bf 71},
062501 (2008), and references therein.

\bibitem{He14} Y. He, Y. Yin, M. Zech, A. Soumyanarayanan, M. M. Yee, T. Williams, M. C.
Boyer, K. Chatterjee, W. D. Wise, I. Zeljkovic, T. Kondo, T. Takeuchi, H. Ikuta, P. Mistark,
R. S. Markiewicz, A. Bansil, S. Sachdev, E. W. Hudson, and J. E. Hoffman, Fermi surface and
 pseudogap evolution in a cuprate superconductor, Science {\bf 344}, 608 (2014).

\bibitem{Chatterjee06} U. Chatterjee, M. Shi, A. Kaminski, A. Kanigel, H. M. Fretwell, K.
Terashima, T. Takahashi, S. Rosenkranz, Z. Z. Li, H. Raffy, A. Santander-Syro, K. Kadowaki,
M. R. Norman, M. Randeria, and J. C. Campuzano, Nondispersive Fermi arcs and the absence of
charge ordering in the pseudogap phase of Bi$_{2}$Sr$_{2}$CaCu$_{2}$O$_{8+\delta}$, Phys.
Rev. Lett. {\bf 96}, 107006 (2006).

\bibitem{Comin16} See, e.g., the review, Riccardo Comin and Andrea Damascelli, Resonant
X-ray scattering studies of charge order in cuprates, Annu. Rev.Condens. Matter Phys.
{\bf 7}, 369 (2016).

\bibitem{Yin21} See, e.g., the review, J.-X. Yin, S. H. Pan, M. Z. Hasan, Probing
topological quantum matter with scanning tunnelling microscopy, Nat. Rev. Phys.
{\bf 3}, 249 (2021).

\bibitem{Pan01} S. H. Pan, J. P. \'ONeal, R. L. Badzey, C. Chamon, H. Ding, J. R. 
Engelbrecht, Z. Wang, H. Eisaki, S. Uchida, A. K. Gupta, K.-W. Ng, E. W. Hudson, K. M. 
Lang, and J. C. Davis, Microscopic electronic inhomogeneity in the high-$T_c$ 
superconductor Bi$_{2}$Sr$_{2}$CaCu$_{2}$O$_{8+x}$, Nature {\bf 413}, 282 (2001).

\bibitem{Fujita19} S. Mukhopadhyay, R. Sharma, C. K. Kim, S. D. Edkins, M. H. Hamidian, H. 
Eisaki, S. Uchida, E.-A. Kim, M. J. Lawler, A. P. Mackenzie, J. C. S. Davis, and K. Fujita, 
Evidence for a vestigial nematic state in the cuprate pseudogap phase, Proc. Natl. Acad. 
Sci. {\bf 116}, 13249 (2019).

\bibitem{Dessau91} D. S. Dessau, B. O. Wells, Z.-X. Shen, W. E. Spicer, A. J. Arko, R. S. 
List, D. B. Mitzi, and A. Kapitulnik, Anomalous spectral weight transfer at the 
superconducting transition of Bi$_{2}$Sr$_{2}$CaCu$_{2}$O$_{8+\delta}$, Phys. Rev. Lett. 
{\bf 66}, 2160 (1991).

\bibitem{Norman97} M. R. Norman, H. Ding, J. C. Campuzano, T. Takeuchi, M. Randeria, T. 
Yokoya, T. Takahashi, T. Mochiku, and K. Kadowaki, Unusual dispersion and line shape of 
the superconducting state spectra of Bi$_{2}$Sr$_{2}$CaCu$_{2}$O$_{8+\delta}$, Phys. Rev. 
Lett. {\bf 79}, 3506 (1997).

\bibitem{Campuzano99} J. C. Campuzano, H. Ding, M. R. Norman, H. M. Fretwell, M. Randeria, 
A. Kaminski, J. Mesot, T. Takeuchi, T. Sato, T. Yokoya, T. Takahashi, T. Mochiku, K. 
Kadowaki, P. Guptasarma, D. G. Hinks, Z. Konstantinovic, Z. Z. Li, and H. Raffy, Electronic 
spectra and their relation to the ($\pi,\pi$) collective mode in high-$T_c$ superconductors, 
Phys. Rev. Lett. {\bf 83}, 3709 (1999).

\bibitem{Wei08} J. Wei, Y. Zhang, H. W. Ou, B. P. Xie, D. W. Shen, J. F. Zhao, L. X. Yang, 
M. Arita, K. Shimada, H. Namatame, M. Taniguchi, Y. Yoshida, H. Eisaki, and D. L. Feng, 
Superconducting coherence peak in the electronic excitations of a single-layer 
Bi$_{2}$Sr$_{1.6}$La$_{0.4}$CuO$_{6+\delta}$ cuprate superconductor, Phys. Rev. Lett. 
{\bf 101}, 097005 (2008).

\bibitem{DMou17} Daixiang Mou, Adam Kaminski, and Genda Gu, Direct observation of self-energy
signatures of the resonant collective mode in Bi$_{2}$Sr$_{2}$CaCu$_{2}$O$_{8+\delta}$, Phys.
Rev. B {\bf 95}, 174501 (2017).

\bibitem{Kaminski01} A. Kaminski, M. Randeria, J. C. Campuzano, M. R. Norman, H. Fretwell, J.
Mesot, T. Sato, T. Takahashi, and K. Kadowaki, Renormalization of spectral line shape and
dispersion below $T_c$ in Bi$_{2}$Sr$_{2}$CaCu$_{2}$O$_{8+\delta}$, Phys. Rev. Lett. {\bf 86},
1070 (2001).

\bibitem{Zhou03} X. J. Zhou, T. Yoshida, A. Lanzara, P. V. Bogdanov, S. A. Kellar, K. M. Shen,
W. L. Yang, F. Ronning, T. Sasagawa, T. Kakeshita, T. Noda, H. Eisaki, S. Uchida, C. T. Lin,
F. Zhou, J. W. Xiong, W. X. Ti, Z. X. Zhao, A. Fujimori, Z. Hussain, and Z.-X. Shen, Universal
nodal Fermi velocity, Nature {\bf 423}, 398 (2003).

\bibitem{Anzai10} H. Anzai, A. Ino, T. Kamo, T. Fujita, M. Arita, H. Namatame, M. Taniguchi,
A. Fujimori, Z.-X. Shen, M. Ishikado, and S. Uchida, Energy-dependent enhancement of the
electron-coupling spectrum of the underdoped Bi$_{2}$Sr$_{2}$CaCu$_{2}$O$_{8+\delta}$ 
superconductor, Phys. Rev. Lett. {\bf 105}, 227002 (2010).

\bibitem{He13} J. He, W. Zhang, J. M. Bok, D. Mou, L. Zhao, Y. Peng, S. He, G. Liu, X. Dong,
J. Zhang, J. S. Wen, Z. J. Xu, G. D. Gu, X. Wang, Q. Peng, Z. Wang, S. Zhang, F. Yang, C.
Chen, Z. Xu, H.-Y. Choi, C. M. Varma, and X. J. Zhou, Coexistence of two sharp-mode couplings
and their unusual momentum dependence in the superconducting state of 
Bi$_{2}$Sr$_{2}$CaCu$_{2}$O$_{8+\delta}$ revealed by laser-based angle-resolved photoemission, 
Phys. Rev. Lett. {\bf 111}, 107005 (2013).

\bibitem{Yang19} S.-L. Yang, J. A. Sobota, Y. He, D. Leuenberger, H. Soifer, H. Eisaki,
P. S. Kirchmann, and Z.-X. Shen, Mode-selective coupling of coherent phonons to the Bi2212
electronic band structure, Phys. Rev. Lett. {\bf 122}, 176403 (2019).

\bibitem{Dessau93} D. S. Dessau, Z. X. Shen, D. M. King, D. S. Marshall, L. W. Lombardo, P. H.
Dickinson, A. G. Loeser, J. DiCarlo, C. H. Park, A. Kapitulnik, and W. E. Spicer, Key features
in the measured band structure of Bi$_{2}$Sr$_{2}$CaCu$_{2}$O$_{8+\delta}$ flat bands at $E_F$
and Fermi surface nesting, Phys. Rev. Lett. {\bf 71}, 2781 (1993).

\bibitem{Eschrig00} M. Eschrig and M. R. Norman, Neutron resonance: modeling photoemission and
tunneling data in the superconducting state of Bi$_{2}$Sr$_{2}$CaCu$_{2}$O$_{8+\delta}$, Phys.
Rev. Lett. {\bf 85}, 3261 (2000).

\bibitem{Manske01} D. Manske, I. Eremin, and K. H. Bennemann, Analysis of the elementary
excitations in high-$T_c$ cuprates: explanation of the new energy scale observed by
angle-resolved photoemission spectroscopy, Phys. Rev. Lett. {\bf 87}, 177005 (2001).

\bibitem{Eschrig02} M. Eschrig and M. R. Norman, Dispersion anomalies in bilayer cuprates
and the odd symmetry of the magnetic resonance, Phys. Rev. Lett. {\bf 89}, 277005 (2002).

\bibitem{Eschrig06} See, e.g., the review, M. Eschrig, The effect of collective spin-1
excitations on electronic spectra in high-$T_c$ superconductors, Adv. Phys. {\bf 55},
47 (2006).

\bibitem{Mahajan13} R. Mahajan, M. Barkeshli, and S. A. Hartnoll, Non-Fermi liquids and
the Wiedemann-Franz law, Phys. Rev. B {\bf 88}, 125107 (2013).

\bibitem{Hartnoll14} S. A. Hartnoll, R. Mahajan, M. Punk, and S. Sachdev, Transport near
the Ising-nematic quantum critical point of metals in two dimensions, Phys. Rev. B
{\bf 89}, 155130 (2014).

\bibitem{Patel14} A. A. Patel and S. Sachdev, dc resistivity at the onset of spin density
wave order in two-dimensional metals, Phys. Rev. B {\bf 90}, 165146 (2014).

\bibitem{Lucas15} A. Lucas and S. Sachdev, Memory matrix theory of magnetotransport in
strange metals, Phys. Rev. B {\bf 91}, 195122 (2015).

\bibitem{Vieira20} L. E. Vieira, V. S. de Carvalho, H. Freire, DC resistivity near a nematic
quantum critical point: Effects of weak disorder and acoustic phonons, Ann. Phys. {\bf 419},
168230 (2020).

\bibitem{Mandal21} I. Mandal and H. Freire, Transport in the non-Fermi liquid phase of
isotropic Luttinger semimetals, Phys. Rev. B {\bf 103}, 195116 (2021).

\bibitem{Prange64} R. E. Prange and L. P. Kadanoff, Transport theory for electron-phonon
interactions in metals, Phys. Rev. {\bf 134}, A566 (1964).

\bibitem{Kuang15} L. Kuang, Y. Lan, and S. Feng, Dynamical spin response in cuprate
superconductors from low-energy to high-energy, J. Magn. Magn. Mater. {\bf 374}, 624 (2015).

\bibitem{Yuan01} F. Yuan, S. Feng, Z. B. Su, and L. Yu, Doping and temperature dependence of
incommensurate antiferromagnetism in underdoped lanthanum cuprates, Phys. Rev. B {\bf 64},
224505 (2001).

\bibitem{Feng98} S. Feng and Z. Huang, Universal spin response in copper oxide materials,
Phys. Rev. B {\bf 57}, 10328 (1998).

\bibitem{Bruin13} J. A. N. Bruin, H. Sakair, R. S. Perry, A. P. Mackenzie, Similarity of
scattering rates in metals showing $T$-linear resistivity, Science {\bf 339}, 804 (2013).

\bibitem{Grigera01} S. A. Grigera, R. S. Perry, A. J. Schofield, M. Chiao, S. R. Julian, 
G. G. Lonzarich, S. I. Ikeda, Y. Maeno, A. J. Millis, and A. P. Mackenzie, Magnetic 
field-tuned quantum criticality in the metallic ruthenate Sr$_{3}$Ru$_{2}$O$_{7}$, 
Science {\bf 294}, 329 (2001).

\bibitem{Ma22} X. Ma, M. Zeng, and S. Feng, unpubliahed.

\bibitem{Mou19} Y. Mou, Y. Liu, S. Tan, and S. Feng, Doping and momentum dependence of 
coupling strength in cuprate superconductors, Phil. Mag. {\bf 99}, 2718 (2019).

\end{thebibliography}
\end{document}